\documentclass[11pt,a4paper]{article}
\usepackage{jheppub}

\usepackage{amsmath}
\usepackage{amssymb}  % checkmark
\usepackage{mathrsfs} % mathscr
\usepackage{graphicx}
\usepackage{dcolumn}
\usepackage{upgreek}
\usepackage{color}
\usepackage{url}
\usepackage[toc,page]{appendix}

\usepackage{pdfpages}

\newcommand{\neu}{\tilde{\chi}_1^0}
\newcommand{\Br}{{\cal B}}

\preprint{\begin{flushright} APCTP-Pre2020-028
\end{flushright}}	
\title{Long-lived light neutralinos at Belle~II}

\author[a]{Sourav Dey}
\emailAdd{souravdey@tauex.tau.ac.il}
\affiliation[a]{School of Physics and Astronomy, Tel Aviv University, Tel Aviv 69978, Israel}

\author[b]{Claudio O. Dib}
\emailAdd{claudio.dib@usm.cl}
\affiliation[b]{Departmento de F\'isica and CCTVal, Universidad T\'ecnica Federico Santa Mar\'ia,\\ Valpara\'iso 2340000, Chile}

\author[c]{Juan Carlos Helo}
\emailAdd{jchelo@userena.cl}
\affiliation[c]{Departamento de F\'{i}sica, Facultad de Ciencias, Universidad de La Serena,\\
Avenida Cisternas 1200, La Serena, Chile}

\author[a]{Minakshi Nayak}
\emailAdd{minakshi@tauex.tau.ac.il}

\author[d]{Nicol\'as A. Neill}
\emailAdd{nicolas.neill@gmail.com}
\affiliation[d]{Instituto de Alta Investigaci\'{o}n, Universidad de Tarapac\'{a}, Casilla 7D, Arica, Chile}

\author[a]{Abner Soffer}
\emailAdd{asoffer@tau.ac.il}

\author[e,f]{Zeren Simon Wang}
\emailAdd{wzs@mx.nthu.edu.tw}
\affiliation[e]{Department of Physics, National Tsing Hua University, Hsinchu 300, Taiwan}
\affiliation[f]{Asia Pacific Center for Theoretical Physics (APCTP) - Headquarters San 31,\\ Hyoja-dong, Nam-gu, Pohang 790-784, Korea}

\abstract{
We consider light neutralinos of mass about 1 GeV, produced from $\tau$ lepton rare decays at Belle~II, in the context of R-parity-violating (RPV) supersymmetry.
With large and clean samples of $\tau$ leptons produced at the Belle~II experiment, excellent sensitivity to such light neutralinos with the exotic signatures of displaced vertices is expected.
We focus on two benchmark scenarios of single RPV operators, $\lambda'_{311} L_3 Q_1 \bar{D}_1$ and $\lambda'_{312} L_3 Q_1 \bar{D}_2$, which induce both the production and decay of the lightest neutralino.
For the reconstruction of a displaced vertex, we require at least two charged pions in the final states.
We perform Monte-Carlo simulations for both signal and background events, and find that Belle~II can explore regions in the parameter space competitive with other probes.
In particular, for the $\lambda'_{311}$ scenario, it can put limits up to two orders of magnitude stronger than the current bounds.
}
\begin{document}

%\keywords{%
%LLP, Belle~II, RPV-SUSY, neutralinos
%}

\maketitle
%%%%%%%%%%%%%%%%%%%%%%%%%%%%%%%%%%%%%%%%%%%%%%%%%%%%%%%%%%%%%%%%%%%%%%

%%%%%%%%%%%%%%%%%%%%%%%%%%%%%%%%%%%%%%%%%%%%%%%%%%%%%%%%%%%%%%%%%%%%%%
\section{Introduction}\label{sec:intro}

Supersymmetry (SUSY) \cite{Nilles:1983ge,Martin:1997ns} remains one of the fore-runners for physics beyond the Standard Model (BSM).
In particular, by predicting superpartners for the standard model (SM) particles, whose contributions to the Higgs boson self-energy cancel those of the SM particles, it solves the hierarchy problem \cite{Gildener:1976ai,Veltman:1980mj} in an elegant manner.
In order to preserve technical naturalness, SUSY predicts such new heavy particles with masses not much higher than the TeV scale.
SUSY searches at the Large Hadron Collider (LHC) at CERN in Geneva, Switzerland, have mostly focused on such heavy particles, which are expected to decay promptly into SM particles with large transverse momentum. 
Furthermore, in the minimal supersymmetric standard model (MSSM), a $Z_2$ symmetry called R-parity is usually assumed in order to avoid proton decay. R-parity conservation (RPC) implies that the lightest supersymmetric particle (LSP) is stable. If the LSP is neutral under quantum chromodynamics (QCD) and quantum electrodynamics (QED), its production in squarks and sleptons decays will lead to large transverse missing energy at LHC. 
However, in SUSY with R-parity-violation (RPV) (see Refs.~\cite{Dreiner:1997uz,Barbier:2004ez,Mohapatra:2015fua} for reviews), one can still circumvent the phenomenological issue of proton decay by imposing a different discrete symmetry e.g., the baryon triality $B_3$ symmetry \cite{Ibanez:1991pr,Dreiner:2012ae}.
Such RPV-SUSY models are equally legitimate and in fact offer a rich phenomenology at colliders. In particular, with RPV, the LSP is no longer stable and can decay into SM particles.

In either the RPC or RPV scenarios, the LHC has so far not discovered SUSY particles of any kind, but only placed TeV-scale lower bounds on the masses of the predicted squarks and gluinos~\cite{Aaboud:2018doq,Sirunyan:2017nyt,Sirunyan:2019mbp,Sirunyan:2019ctn,Aad:2020nyj}. 
This has led to increased interest in other BSM signatures.
One class of signatures of interest involves long-lived particles (LLPs), which are generally defined as particles that travel a macroscopic distance before decaying.
In fact, even in the SM, such LLPs are common, e.g., the pion, kaon, muon, and neutron.
Hence, it is perhaps no surprise that LLPs are predicted in a large number of classes of BSM models. 
Well studied examples include gauge-mediated, RPV, and split SUSY models, neutral-naturalness models, and portal physics that connects the SM and a ``dark sector'', where the scalar/fermion/pseudoscalar/vector portal predicts a dark scalar/heavy neutral lepton/axion-like particle/dark photon.
For recent reviews of LLP searches, see Refs.~\cite{Alimena:2019zri,Lee:2018pag,Curtin:2018mvb}.

In this work, we consider a long-lived, light neutralino LSP within the RPV-SUSY.
Although there are increasingly stronger lower limits on the squark and gluino masses, this is not the case for the lightest neutralino.
In fact, if the GUT (grand-unified theory) relation $M_1= \frac{5}{3}\tan^2{\theta_W}M_2$ between the Bino and Wino masses is lifted and the dark matter constraint is dropped, the neutralino can be as light as in the GeV scale or even massless \cite{Choudhury:1995pj,Choudhury:1999tn,Belanger:2002nr,Bottino:2002ry,Belanger:2003wb,Vasquez:2010ru,Calibbi:2013poa,Gogoladze:2002xp,Dreiner:2009ic}.
Such light neutralinos are also consistent with both astrophysical and cosmological constraints \cite{Grifols:1988fw,Ellis:1988aa,Lau:1993vf,Dreiner:2003wh,Dreiner:2013tja,Profumo:2008yg,Dreiner:2011fp}.
Note that in order to avoid overclosing the Universe \cite{Hooper:2002nq, Bottino:2011xv, Belanger:2013pna, Bechtle:2015nua}, light neutralinos should decay, e.g., in the context of the RPV-SUSY.
Moreover, for a range of values of the RPV couplings, GeV-scale neutralinos are long-lived and have a significant probability of decaying inside a detector. Such decays lead to a displaced-vertex (DV) signature at colliders, which is the signature we explore here.

The GeV-scale neutralinos are necessarily Bino-like in order to avoid the present bounds~\cite{Gogoladze:2002xp,Dreiner:2009ic}.
In the literature, two main production mechanisms have been discussed and studied for the LHC~\cite{deVries:2015mfw}, future lepton colliders~\cite{Wang:2019orr,Wang:2019xvx}, and a list of proposed extended programs at the LHC~\cite{Helo:2018qej,Dercks:2018eua,Dercks:2018wum,Dreiner:2020qbi}: 1) pair production in $Z$-boson rare decays via the small Higgsino component, and 2) single production in rare decays of charm and bottom mesons via an RPV coupling.
In this work, for the first time, we propose to consider light neutralinos produced in $\tau$ lepton decays.

$\tau$ leptons are copiously produced at colliders, including the LHC, $B$-factories, and $\tau$-charm threshold colliders~\cite{Achasov:2019rdp}.
Here we focus on the ongoing Belle~II experiment in Japan~\cite{Abe:2010gxa,Kou:2018nap} at the intensity-frontier, where electron and positron beams are colliding at a center-of-mass energy $\sqrt{s} = 10.58$ GeV.
The projected total integrated luminosity of Belle~II, to be collected in the next few years, is 50~ab$^{-1}$, roughly 50 times that of the previous Belle experiment. This corresponds to a sample of $4.6\times 10^{10}$ $e^+e^-\to\tau^+\tau^-$ events. The resulting events are easily identifiable via the back-to-back production of the $\tau$ pairs and the large missing momentum. These properties make Belle~II one of the best facilities for the study of rare $\tau$ decays.

The LLP search potential of Belle~II and the previous-generation $B$-factories, BABAR~\cite{Aubert:2001tu} and Belle~\cite{Abashian:2000cg}, for a variety of models has been investigated extensively (see for instance Refs.~\cite{Batell:2009yf,Canetti:2014dka,Dolan:2017osp,Sullivan:2017qdj, Cvetic:2017vwl,Dib:2019tuj,Filimonova:2019tuy,Duerr:2019dmv}).
In particular, Ref.~\cite{Dib:2019tuj} considered heavy neutral leptons as the LLPs produced from $\tau$ decays at Belle~II, using as a benchmark model a heavy neutral lepton that mixes predominantly with the third-generation active neutrino $\nu_\tau$.

This paper is structured as follows.
Section~\ref{sec:model} introduces the RPV-SUSY model considered in this work, followed by Sec.~\ref{sec:prodANDdecay} in which we present the analytic formulas of $\tau$ and neutralino decays.
In Sec.~\ref{sec:bgd} we explain the signature definition and background estimate, followed by Sec.~\ref{sec:sensitivity} where we elaborate on the sensitivity estimate procedure via a Monte-Carlo (MC) simulation.
We present the numeric results in Sec.~\ref{sec:results}.
Finally, we summarize the work and offer an outlook in Sec.~\ref{sec:conclu}.

\section{Model basics and possible displaced-vertex signatures}\label{sec:model}

We consider a model where the MSSM is appended with R-parity violation (RPV-MSSM).
With the R-parity violation, the usual MSSM superpotential is extended with the following terms:
\begin{equation}
    W_{\text{RPV}} = \epsilon_i L_i \cdot H_u + \frac{1}{2} \lambda_{ijk} L_i \cdot L_j \bar E_k + \lambda'_{ijk} L_i \cdot Q_j \bar D_k + 
    \frac{1}{2}\lambda''_{ijk}\bar U_i \bar D_j \bar D_k ,\label{eq:rpvlag}
\end{equation}
where the indices $i,j,k$ refer to any of the three fermion generations. The first three sets of terms are lepton-number-violating, and the last set violates baryon number.  
Allowing all the terms would lead to a too large proton decay rate, unless the couplings were extremely small.
As mentioned above, imposing certain discrete symmetries, such as the baryon triality, can remove, e.g., the baryon-number-violating terms, thus forbidding proton decay.
For simplicity, we assume that only certain $\lambda'_{ijk} L_i Q_j \bar{D}_k$ operators are non-vanishing.
More concretely, we focus on the operators $\lambda'_{311} L_3 Q_1 \bar{D}_1$ and $\lambda'_{312} L_3 Q_1 \bar{D}_2$, which are relevant for $\tau$ decay.

In the RPV-MSSM, while the decay of the lightest neutralino can only proceed via one or more RPV couplings, its production can be mediated via different mechanisms.
First, despite the Bino-like nature of such light neutralinos, the small Higgsino component allows for coupling with a $Z$-boson.
If $\neu$ is lighter than $M_Z/2$, the decay $Z\rightarrow \neu \neu$ may ensue.
This scenario has been studied in the context of both LHC experiments and future $Z$-factories.
Second, the $LQ\bar{D}$ operators can lead to the decay of charm and bottom mesons into the lightest neutralino.
This was proposed for the first time in Ref.~\cite{deVries:2015mfw}.
Providing the analytic expressions of the decay widths of the mesons and light neutralinos, the authors studied the sensitivity of ATLAS and the fixed-target experiment SHiP \cite{SHiP:2018yqc}.
It is worth mentioning that the authors of Ref.~\cite{deVries:2015mfw} estimated the transition form factors for both the second and third generations of quarks using the heavy quark formulation.
While they cannot rigorously apply such a method to purely first-generation flavors, they state that it is reasonable to follow the same approach and to assume that the tensor and vector decay constants satisfy $f_T = f_V$ for first-generation mesons as well, within the expected level of precision determined by other uncertainties. 
In our benchmark scenarios, which are dominated by  pseudoscalar  mesons, we use the same assumption because the impact of the uncertainty in $f_T$ is even smaller.

\begin{figure}[!htbp]
\includegraphics[scale=0.8]{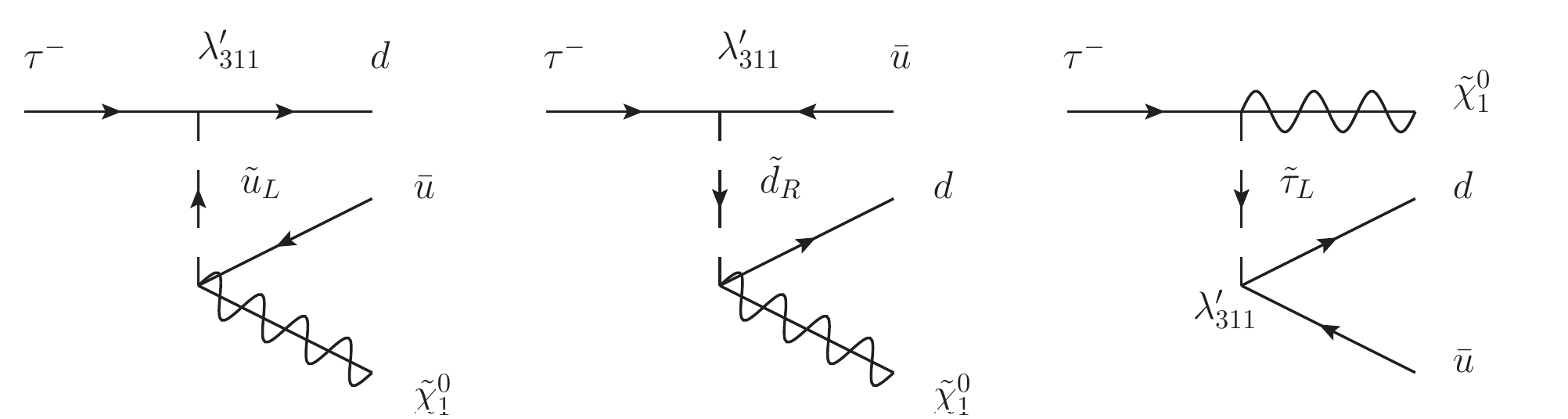}
\includegraphics[scale=0.8]{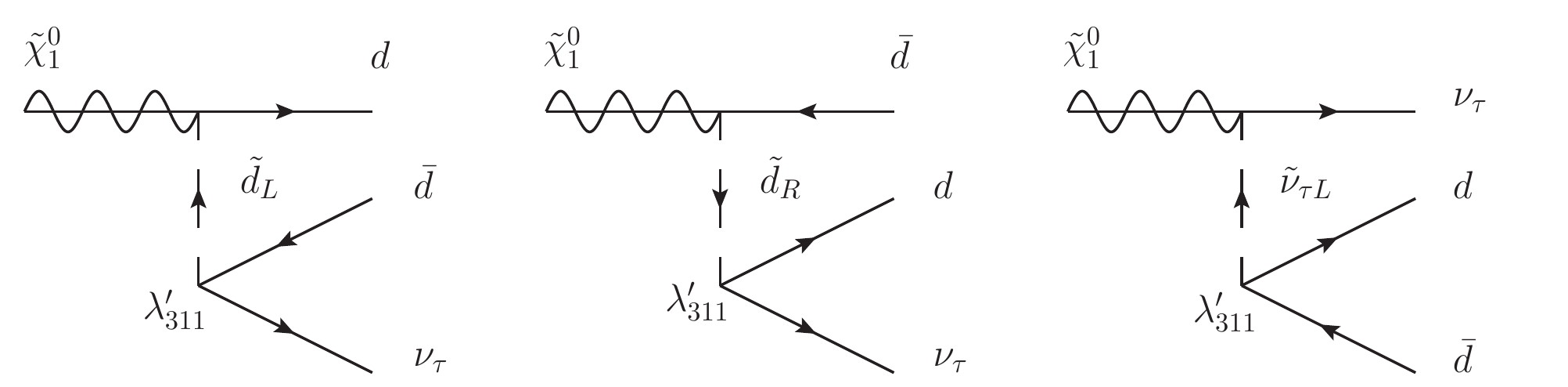}
\caption{The parton-level Feynman diagrams for both $\tau$ and $\neu$ decays via the coupling $\lambda'_{311}$. The diagrams for the coupling $\lambda'_{312}$ can be easily obtained by changing a $d$ (s)quark to an $s$ (s)quark.
}
\label{fig:feynman_diag}
\end{figure}

In this paper we consider the case of one single RPV operator, either $\lambda'_{311} L_3 Q_1 \bar{D}_1$ only or $\lambda'_{312} L_3 Q_1 \bar{D}_2$ only, mediating both the production and decay of a light neutralino.
As shown in Fig.~\ref{fig:feynman_diag}, the $\tau^-$ decay produces a charged pseudoscalar (vector) meson $M_1$ ($M_1^*$), with quark content $d\bar u$ (namely, $\pi^-$, $\rho^-$) in the case of $\lambda'_{311}\ne 0$, or $s\bar u$ ($K^-$, $K^{*-}$) in the case of $\lambda'_{312}\ne 0$.
The same $LQ\bar D$ operator also gives rise to the neutralino decay.
Since the neutralino is produced from $\tau$ decays, its mass satisfies $m_{\neu} < m_\tau$.
As a result, it can only decay into $\nu_\tau + M_2^{(*)}$, where $M_2^{(*)}$ is a pseudoscalar (vector) neutral meson with quark content 
$d\bar{d}$ (forming either a $\pi^0$, $\rho^0$, $\eta$, $\eta'$, or $\omega$) for the $\lambda'_{311}$ coupling, or $d\bar{s}$ (forming a $K^0$, $K^{*0}$) for the $\lambda'_{312}$ coupling (see Fig.~\ref{fig:feynman_diag}).
Except for the $\pi^0$ and $K^0$, these mesons decay promptly into charged hadrons, enabling the identification of the neutralino displaced decay position.
When $M_2$ is a $K^0$ that decays as $K_S\to\pi^+\pi^-$, the DV does not identify the decay position of the neutralino, because of the long lifetime of the $K_S$ ($c\tau_{K_S}\approx 2.7$~cm).
Nonetheless, this signature can be used to detect a signal if the neutralino flight distance is significantly larger than $(p_{K_S}/m_{K_S})c\tau_{K_S}$ and the momentum vector of the $K_S$ does not point back to the $\tau$ decay position.
Owing to the relatively short lifetime of the $\tau$, its decay position can be safely taken to be the interaction point (IP) of the collider beams.

%%%%%%%%%%%%%%%%%%

The current bound on $\lambda'_{31k}$ is calculated in Ref.~\cite{Bansal:2019zak} for the case in which the virtual sfermion is a squark with mass $m_{\tilde{q}_R}\gtrsim1$~TeV:
\begin{eqnarray}
|\lambda'_{31k}| < 0.20 \, \frac{m_{\tilde{d}_{Rk}}}{1\text{ TeV}} + 0.046. \label{eq:rpvlimit31k}
\end{eqnarray}
This limit was obtained by recasting a $pp\to W' \to \tau\nu_\tau$ search for high-mass resonances decaying to a tau lepton and a neutrino, performed at ATLAS with 36.1~fb$^{-1}$ of data~\cite{Aaboud:2018vgh}. A similar search was performed by CMS \cite{1807.11421}.
The expected bound at the HL-LHC with 3~ab$^{-1}$ is also given in Ref.~\cite{Bansal:2019zak}. Since it is only slightly tighter than the present bound of Eq.~(\ref{eq:rpvlimit31k}), we use only the present bound for comparison to our results below.

An additional bound comes from the current uncertainties in the measured branching fractions for $\tau \to P \nu_\tau$ ($P=\pi,K$) \cite{Zyla:2020zbs} and the uncertainty in their theoretical predictions \cite{Decker:1994kw},
\begin{align}
    \mathcal B (\tau \to \pi \nu_\tau)_{EXP} = (10.82\pm 0.05)\%,\ \ \ 
    \mathcal B (\tau \to K \nu_\tau)_{EXP} = (0.696\pm 0.01)\%,\\
    \mathcal B (\tau \to \pi \nu_\tau)_{SM} = (10.90\pm 0.027)\%,\ \ \ 
    \mathcal B (\tau \to K \nu_\tau)_{SM} = (0.722\pm 0.004)\%,
\end{align}
where for the SM values, quoted as a function of the $\tau$ lifetime in Ref.~\cite{Decker:1994kw}, we use the current PDG average value $\tau_\tau = (290.3\pm 0.5) \mbox{ fs}$.
From the combined experimental and theoretical uncertainties in $\mathcal B(\tau\to P\nu_\tau)$,
\begin{equation}
\sigma_{\mathcal B(\tau\to \pi \nu_\tau)} = 0.057\times 10^{-2},\ \ \
\sigma_{\mathcal B(\tau\to K \nu_\tau)} = 0.011\times 10^{-2},
\end{equation}
we extract the following 95\% confidence level bounds on $\mathcal B(\tau \to P \neu)$
\begin{equation}
    \mathcal B(\tau \to P \neu) \lesssim 2 \sigma_{\mathcal B (\tau \to P \nu_\tau)}.
    \label{eq:tauBRlimit}
\end{equation}
This limit is relevant only for large neutralino lifetimes. For short lifetimes, the neutralino daughter particles are in principle visible in the detector and might be rejected from the $\tau \to P \neu$ candidate sample, depending on the analysis criteria. Therefore, we show this bound with our results in Sec.~\ref{sec:results} only for $c\tau_{\neu}>1$~m.
We ignore the fact that this bound becomes inaccurate when $m_{\neu}$ approaches $m_\tau$, so that the the $\tau \to P \neu$ event selection may reject the resulting soft pseudoscalar.

\section{Neutralino production and decay}\label{sec:prodANDdecay}

We extract the effective interaction Lagrangian from Ref.~\cite{deVries:2015mfw},
\begin{align}
\mathcal L = &
\ G_{ijk}^{S,\nu}(\overline{\widetilde{\chi}^0} P_L  \nu_i)  (\overline{d_k} P_L  d_j)
+
G_{ijk}^{S,\ell}(\overline{\widetilde{\chi}^0} P_L  \ell_i) (\overline{d_k} P_L  u_j)\nonumber\\
& +
G_{ijk}^{T,\nu}(\overline{\widetilde{\chi}^0} \sigma^{\mu \nu}  \nu_i) (\overline{d_k} \sigma^{\rho \sigma}  d_j) \Big( g_{\mu \rho} g_{\nu \sigma} - \frac{i \epsilon_{\mu \nu
\rho \sigma}}{2}\Big)\nonumber\\
& +
G_{ijk}^{T,\ell} (\overline{\widetilde{\chi}^0} \sigma^{\mu \nu}  \ell_i) (\overline{d_k} \sigma^{\rho \sigma}  u_j) \Big( g_{\mu \rho} g_{\nu \sigma} - \frac{i \epsilon_{\mu \nu
\rho \sigma}}{2}\Big)+\mbox{h.c.},
\label{eq:effL}
\end{align}
%%%%%%%%%
where the effective couplings $G^{S,T}_{ijk}$ are proportional to the corresponding $\lambda'_{ijk}$ couplings,
\begin{align}
G^{S,\ell}_{ijk} = \lambda'_{ijk}
\left(
  \frac{1}{2}\frac{g_{\tilde u_L}}{m^2_{\tilde u_{jL}}}
 +\frac{1}{2}\frac{g_{\tilde d_R}^*}{m^2_{\tilde d_{kR}}}
 - \frac{g_{\tilde \ell_L}}{m^2_{\tilde \ell_{iL}}}
\right),\ \ \
G^{T,\nu}_{ijk} = \lambda'_{ijk}
\left(
  \frac{g_{\tilde d_L}}{4 m^2_{\tilde d_{jL}}}
 +\frac{g_{\tilde d_R}^*}{4 m^2_{\tilde d_{kR}}}
\right),\\
G^{S,\nu}_{ijk} = \lambda'_{ijk}
\left(
  \frac{g_{\tilde \nu_L}}{m^2_{\tilde \nu_{i L}}}
 -\frac{1}{2}\frac{g_{\tilde d_L}}{m^2_{\tilde d_{jL}}}
 -\frac{1}{2}\frac{g_{\tilde d_R}^*}{m^2_{\tilde d_{kR}}}
\right),\ \ \
G^{T,\ell}_{ijk} = \lambda'_{ijk}
\left(
  \frac{g_{\tilde u_L}}{4 m^2_{\tilde u_{jL}}}
 +\frac{g_{\tilde d_R}^*}{4 m^2_{\tilde d_{kR}}}
\right),\label{eq:GST}
\end{align}
and the coupling constants $g_{\tilde f}$ are given by the electroweak $SU(2)$ gauge coupling $g_2$ and the electroweak mixing angle $\theta_W$:
\begin{align}
g_{\tilde \ell_L} = g_{\tilde \nu_L} = \frac{g_2}{\sqrt{2}}\tan\theta_W,
\ \ \ g_{\tilde u_L} = g_{\tilde d_L} =- \frac{g_2}{3 \sqrt{2}}\tan\theta_W,
\ \ \ g_{\tilde d_R} = - \frac{2 g_2}{3 \sqrt{2}}\tan\theta_W.
\end{align}
%%%%%%%%%
From the effective Lagrangian, the neutralino production rate in association with a scalar meson $M_1$ or a vector meson $M^*_1$ is
\begin{align}
\Gamma(\tau\to M_1 \neu) & = \frac{\lambda^{1/2}(m_\tau^2,m_{M_1}^2,m_{\neu}^2)}{128 \pi m_\tau^3} |G^{S,\ell}_{3jk}|^2 |f_{M_1}^S|^2
(m_\tau^2 - m_{M}^2 +  m_{\neu}^2),\\
\Gamma(\tau\to M^*_1 \neu) =& \frac{\lambda^{1/2}(m_\tau^2,m_{M_1^*}^2,m_{\neu}^2)}{2\pi m_\tau^3} |G^{T,\ell}_{3jk}|^2 |f^T_{M_1^*}|^2\nonumber\\
& \left[2 (m_\tau^2 - m_{\neu}^2)^2 - m_{M_1^*}^2 (m_{M_1^*}^2 + m_\tau^2 + m_{\neu}^2)\right],
\end{align}
where the scalar and tensor decay constants $f_{M_1}^S$, $f_{M_1^*}^T$ are given in Appendix~\ref{app:constants}.
For the decays of the neutralino in our scenario, we reproduce the expressions given  in Ref.~\cite{deVries:2015mfw},
\begin{align}
\Gamma(\neu\to M_2 \nu_{\tau}) & = \frac{\lambda^{1/2}(m_{\neu}^2,m_{M_2}^2,0)}{128 \pi m_{\neu}^3} |G^{S,\nu_\tau}_{3jk}|^2 |f_{M_2}^S|^2
(m_{\neu}^2 - m_{M_2}^2),\\
\Gamma(\neu\to M_2^* \nu_\tau) & = \frac{\lambda^{1/2}(m_{\neu}^2,m_{M_2^*}^2,0)}{2\pi m_{\neu}^3} |G^{T,\nu_\tau}_{3jk}|^2 |f_{M_2^*}^{T}|^2
\left[2 m_{\neu}^4 - m_{M_2^*}^2 (m_{M_2^*}^2 + m_{\neu}^2)\right],
\end{align}
where $\lambda(x,y,z)=x^2+y^2+z^2-2xy-2xz-2yz$. 
Note that the charge-conjugate channel for the $\neu$ decays is implied.
For simplicity, we assume degenerate sfermion masses, henceforth denoted $m_{\tilde f}$, enabling a study of the sensitivity in terms of the ratio $\lambda'/m_{\tilde f}^2$.

\begin{figure}[!htbp]
\centering
\includegraphics[width=0.48\textwidth]{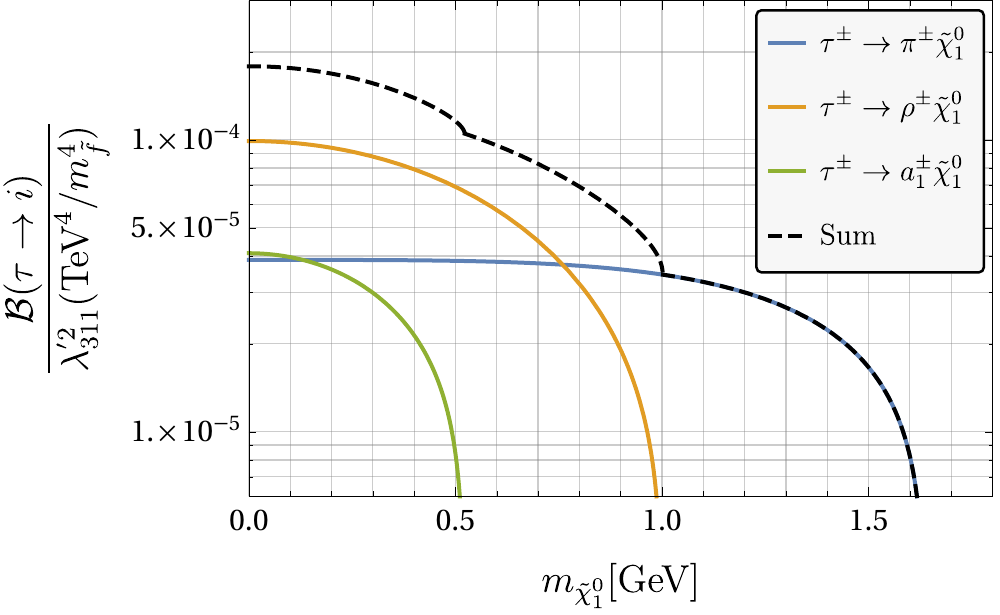}
\includegraphics[width=0.47\textwidth]{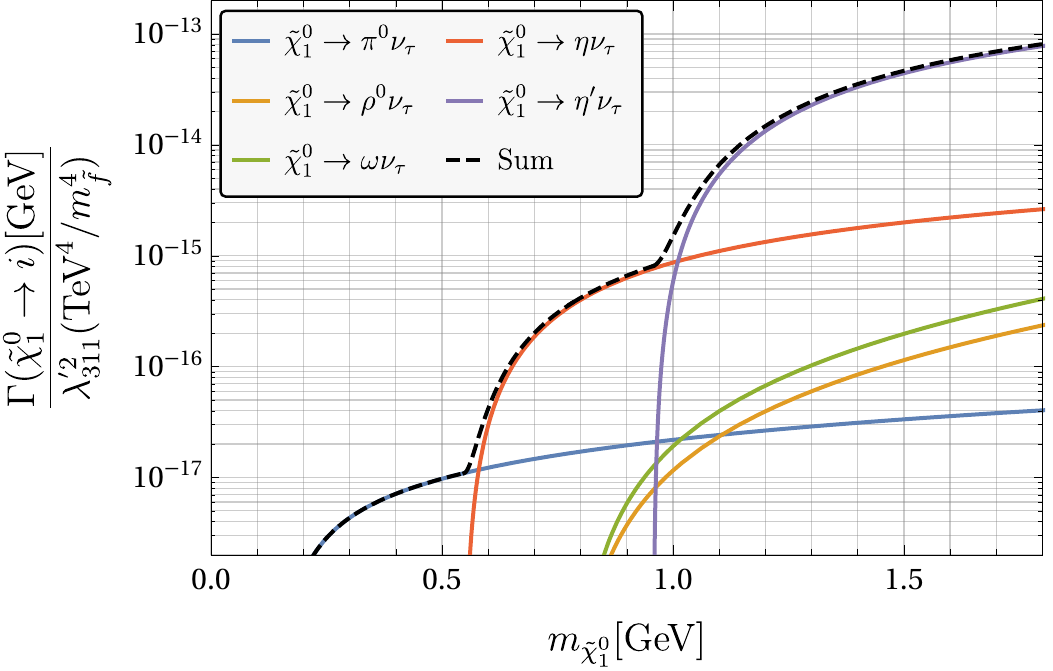}
\includegraphics[width=0.45\textwidth]{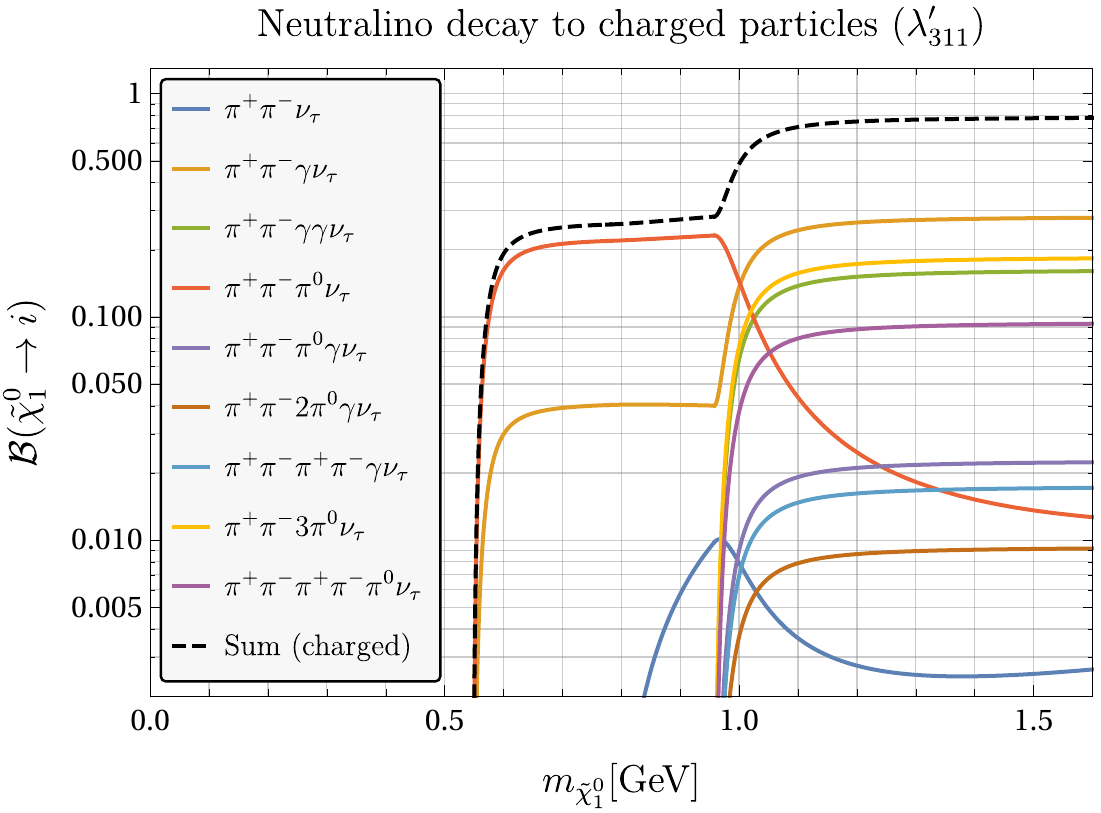}
\caption{Neutralino production and decay in the $\lambda'_{311}\neq 0$ scenario.
(Top left) Branching fractions $\Br(\tau\to M_1\neu)$ for $\tau$ decays into a meson and a neutralino.
(Top right) Decay rates $\Gamma(\neu\to M_2\nu_\tau)$ for neutralino decays into a meson and a neutrino.
(Bottom) Branching fractions for neutralino decays into final states with charged particles, accounting for the decays of the $M_2$ mesons.}
\label{fig:neutralino_production_and_decay_l311}
\end{figure}

\begin{figure}[!htbp]
\centering
\includegraphics[width=0.48\textwidth]{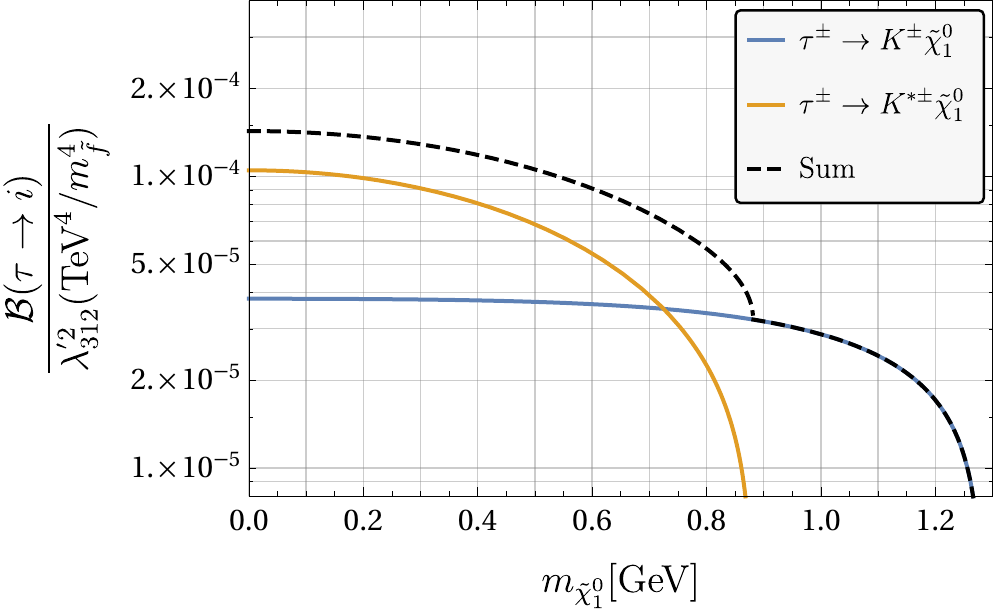}
\includegraphics[width=0.48\textwidth]{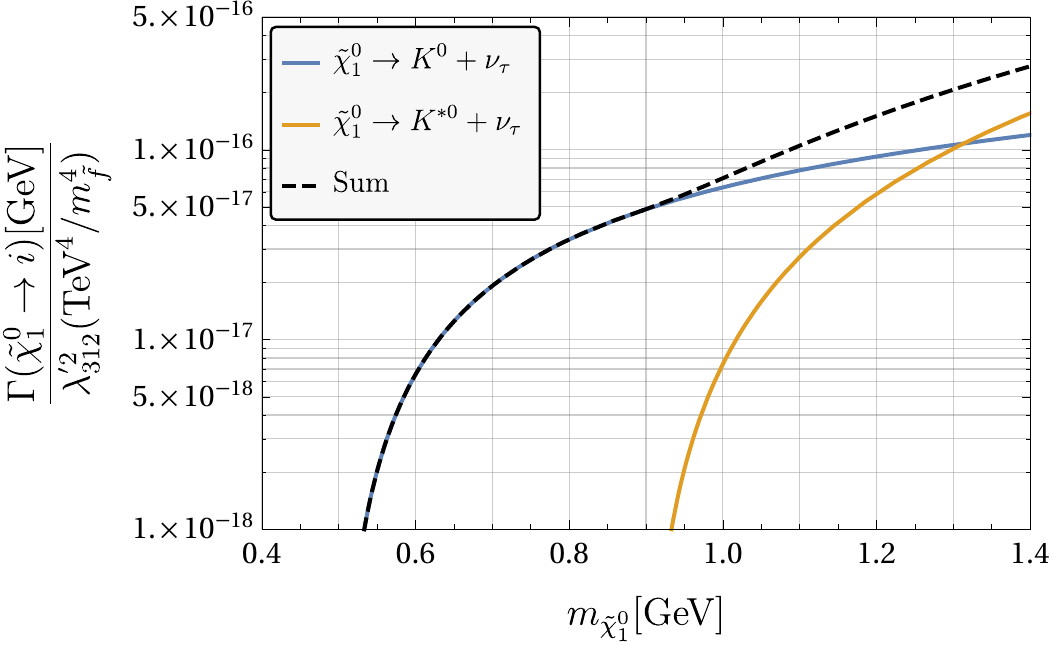}
\includegraphics[width=0.45\textwidth]{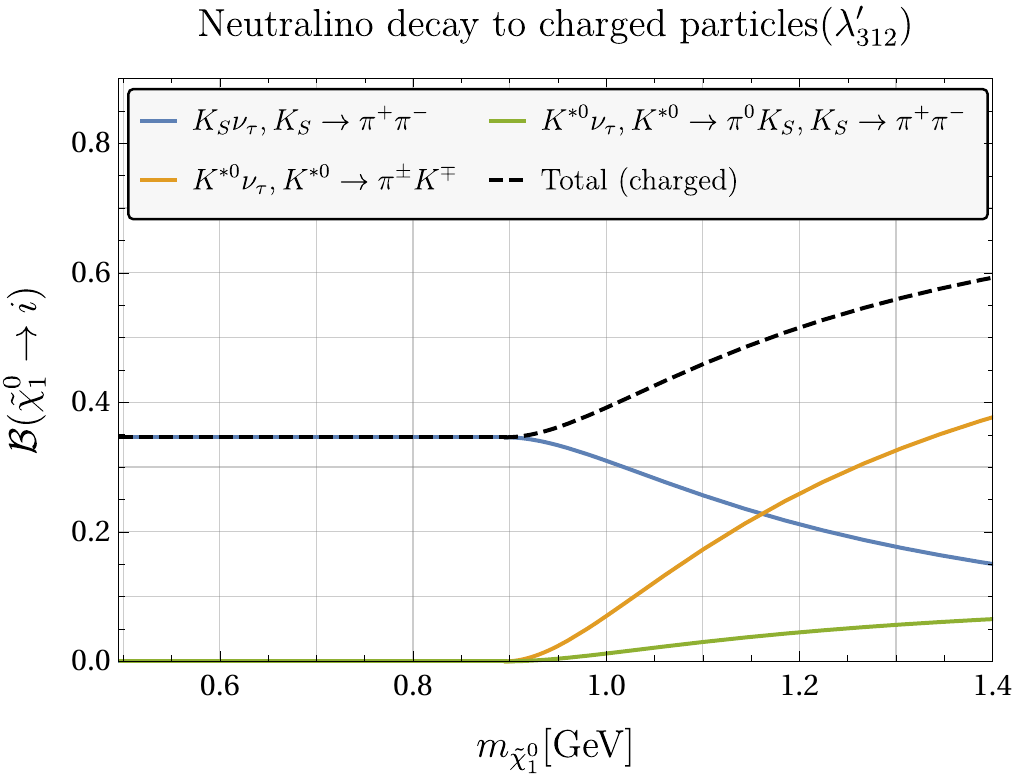}
\caption{Neutralino production and decay in the $\lambda'_{312}\neq 0$ scenario. (Top left) Branching fractions $\Br(\tau\to M_1\neu)$ for $\tau$ decays into a meson and a neutralino.
(Top right) Decay rates $\Gamma(\neu\to M_2\nu_\tau)$ for neutralino decays into a meson and a neutrino.
(Bottom) Branching fractions for neutralino decays into final states with charged particles, accounting for the decays of the $M_2$ mesons.}
\label{fig:neutralino_production_and_decay_l312}
\end{figure}

Table \ref{tab:final-states-summary} lists the different decay modes that we consider for the cases of non-zero $\lambda'_{311}$ or $\lambda'_{312}$.
In Figs.~\ref{fig:neutralino_production_and_decay_l311} and \ref{fig:neutralino_production_and_decay_l312} we show the rates for the different neutralino production and decay modes, as well as the decay branching ratios of the neutralino to visible modes.
It is evident that neutralino decays into pseudoscalar $M_2$ mesons, namely, $\eta$ and $\eta'$, are by far dominant, and hence constitute the best search strategies. For this reason, in the case $\lambda'_{311}\ne 0$ we do not consider neutralino production associated with the $a_1^-$ meson, i.e.  $\tau^-\to \neu a_1^-$. Since $m_\tau - m_{a_1}<m_{\eta}$, this mode does not contribute to visible DVs in this scenario and hence is irrelevant in this estimate.
Also shown in the figures are the neutralino branching fractions into final states with at least two charged pions, which are needed for reconstruction of the DV and its position. 

\begin{table}[!htbp]
	\begin{center}
	\resizebox{\textwidth}{!}{%
		\begin{tabular}{| c || c | c |}
			\hline
			 \multicolumn{3}{|c|}{$\tau \to \neu \  M_1^{(*)}$, $\neu \to M_2^{(*)} \  \nu_\tau$}\\
			\hline
			\hline
			%Scenarios 
			&   Scenario 1 & Scenario 2 \\
			\hline
			$\lambda^\prime$ (production and decay) & $\lambda^\prime_{311}$ & $\lambda^\prime_{312}$\\
			\hline
			Mesons in $\neu$ production ($M_1$) & $\pi^\pm$, $\rho^\pm$ & $K^\pm$ $K^{*\pm}$ \\
			\hline
			Mesons in $\neu$ decay ($M_2$) & $\pi^0$, $\rho$, $\omega$, $\eta$, $\eta'$ &  $K^0$, $K^{*0}$\\
			\hline
			$M_2$ decays with charged &
			$\eta \to \pi^+\pi^-\gamma$, $\eta\to \pi^+\pi^-\pi^0$, & 
			\quad $K_S\to \pi^+\pi^-$,
			\quad { }
			\\
			 particles    &  
			     $\eta'\to \pi^+\pi^-(\eta\to 3\pi^0)$, 
			     $\eta'\to \pi^0\pi^0(\eta\to \pi^+\pi^-\pi^0)$,
			     &  $K^{*0}\to \pi^\pm K^\mp$
			     \\
			 & $\eta'\to \pi^+\pi^-(\eta\to \gamma\gamma)$, $\eta'\to \pi^0\pi^0(\eta\to \pi^+\pi^-\gamma)$  & \\
			 & $\eta'\to \gamma (\omega\to \pi^+\pi^-\pi^0)$, $\eta'\to \gamma (\rho\to \pi^+\pi^-)$& \\
			 & $\eta'\to \gamma (\omega\to \pi^+\pi^-)$, $\eta'\to \pi^+\pi^-(\eta\to \pi^+\pi^-\gamma)$ & \\
			 & $\eta'\to \pi^+\pi^-(\eta\to \pi^+\pi^-\pi^0)$ & \\
			\hline
		\end{tabular}
		}
		\caption{Benchmark scenarios for neutralinos produced from $\tau$ decays: $\tau \to \neu M^{(*)}_1$, $\neu \to M^{(*)}_2 \nu_\tau$.}
		\label{tab:final-states-summary}
	\end{center}
\end{table}

%%%%%%%%%%%%%%%%%
%
%

\section{Event selection and background estimate}\label{sec:bgd}
The search proposed here is experimentally very similar to that of the heavy neutral lepton (HNL) search proposed in Ref.~\cite{Dib:2019tuj}. Therefore, the background suppression and estimation methods, as well as conclusions about the major sources of background, are similar to those of Ref.~\cite{Dib:2019tuj}. In particular, the selection of signal events will begin with the typical $e^+e^-\to \tau^+\tau^-$ selection criteria. These include a single track that recoils against the rest of the event in the opposite hemisphere, accompanied by large missing energy and transverse momentum~\cite{delAmoSanchez:2010pc,Dib:2019tuj}. This selection leaves the backgrounds from $e^+e^-\to \tau^+\tau^-$ events as the dominant source, efficiently rejecting all other sources of background.

Subsequently, the main LLP requirement involves selection of two identified charged pions that originate from a high-quality vertex that is significantly displaced from the IP. Furthermore, a key difference with respect to Ref.~\cite{Dib:2019tuj} is that the neutralino decay produces a narrow hadron, namely, a $\eta$ or $\eta'$, which is rarely produced in $\tau$ decays~\cite{Pich:2013lsa,Nussinov:2009sn,Nussinov:2008gx,delAmoSanchez:2010pc}.
Requiring the presence of a $\eta$ or $\eta'$ decay that occurs far from the interaction point will be very effective at suppressing the background. 
In what follows we estimate the background for the signal case $\tau^-\to \pi^-\neu$, $\neu\to\eta\nu_\tau$, $\eta\to \pi^+\pi^-\pi^0$ in detail.
We then qualitatively extend our conclusions to the other signal modes.

Na\"\i vely, since the $\eta$ is reconstructed in the final state $\pi^+\pi^-\pi^0$,
a conceivable source of background is $\tau^-\to \pi^- K_L \nu_\tau$, followed by the decay of the long-lived $K_L$ to $\pi^+\pi^-\pi^0$.
The branching fractions of these two decays are $4.2\times 10^{-3}$ and 12.5\%, respectively. The $\pi^+\pi^-\pi^0$ mass resolution is less than 4~MeV at $B$-factories~\cite{delAmoSanchez:2010pc}.
Therefore, even if one accounts for the resolution degradation in the case of a displaced decay, the likelihood of a $K_L$ decay faking an $\eta$ is greatly suppressed, given that the masses of these mesons differ by about 50~MeV. 

We estimate that the largest source of background will arise from $\tau^-\to \pi^-\pi^0_p K_L \nu_\tau$, which has a branching fraction of $1.7\times 10^{-3}$, followed by $K_L\to \pi^+\pi^-\pi^0_d$, for which the branching fraction is 12.5\%. For clarity, we denote the prompt and displaced $\pi^0$ mesons with the subscripts $p$ and $d$, respectively. The photons that are produced in the $\pi^0_p$ and $\pi^0_d$ decays are denoted $\gamma_p^i$ and $\gamma_d^i$, with $i=1,2$. A fake $\eta\to \pi^+\pi^-\pi^0$ decay candidate can then be reconstructed with the displaced $\pi^\pm$ and either the $\pi^0_p$ or a fake $\pi^0$ candidate formed from $\gamma_p^i \gamma_d^j$. The two remaining photons would, if detected, indicate that this is not a signal event. However, they may escape via the endcap openings in the calorimeter or otherwise go undetected, typically since they are too soft to be clearly identified above photon background that originates from the collider itself.

To estimate the level of this background, we use the EvtGen~\cite{Lange:2001uf} Monte-Carlo generator to produce $e^+e^-\to \tau^+\tau^-$ where one of the $\tau$ leptons decays via this background decay chain. No detector simulation is used. Following Ref.~\cite{Dib:2019tuj}, we require the $K_L$ decay position to be in the fiducial volume defined by the radial and longitudinal ranges $10<r<80$~cm, $-40<z<120$~cm. The background events have an acceptance of 15\% relative to this selection, which is motivated by the long lifetime of the neutralino and by the desire to suppress background from prompt tracks and from material-interaction background. 

We form a $\pi^0$ candidate from two photons other than $\gamma_d^1 \gamma^2_d$. Each photon is required to have a lab-frame energy of at least 100~MeV. This requirement is tighter than the usual selection in $\tau$ decays~\cite{delAmoSanchez:2010pc}, and thus leads to a more robust and conservative background estimate.
The Belle~II crystal calorimeter~\cite{Abe:2010gxa} has only limited ability to determine the flight direction of a photon, and we conservatively ignore this ability altogether. Thus, we take the direction of the photon momentum from its hit position on the surface of the calorimeter, assuming that it originated from the $K_L$ decay position, which will be experimentally known from the $\pi^+\pi^-$ decay vertex. 
Following Ref.~\cite{delAmoSanchez:2010pc}, we require the invariant mass of the $\pi^0$ candidate to be within 15~MeV of the known value~\cite{Zyla:2020zbs}. Similarly, the invariant mass of the $\eta$ candidate formed from the $\pi^0$ candidate and the displaced $\pi^+\pi^-$ pair is required to be within 15~MeV of the known value, corresponding to about 3 times the resolution. Since the $\gamma^1_p \gamma^2_d$ and the $\pi^+\pi^-\gamma^1 \gamma^2$ mass spectra are broad and smooth, the lack of a detector simulation does not significantly impact our rough estimate. We find that these requirements retain 2.5\% of the background events that are within the fiducial region.  
 
We further require that each of the additional photons in the event be unobservable with conservative requirements, namely, escape the detector through the endcap openings in the calorimeter~\cite{Abe:2010gxa} or have a lab-frame energy smaller than 100~MeV. This requirement retains 3.9\% of the background events selected up to this point. As a whole, we expect a total of about 300 background events in the Belle~II dataset.

\begin{figure}[!htbp]
    \centering
    \begin{tabular}{cc}
    \includegraphics[width=0.5\textwidth]{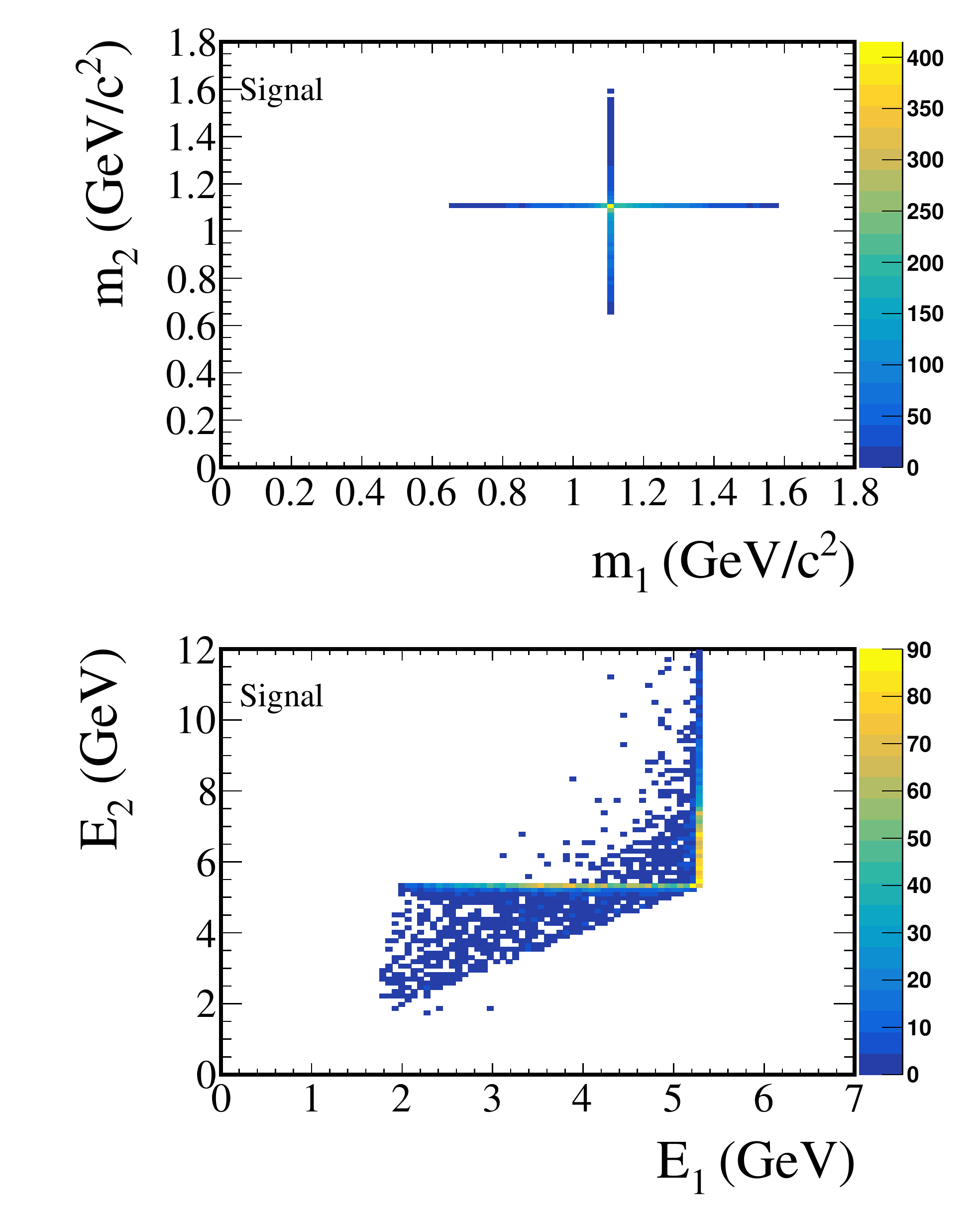}
    &
    \includegraphics[width=0.5\textwidth]{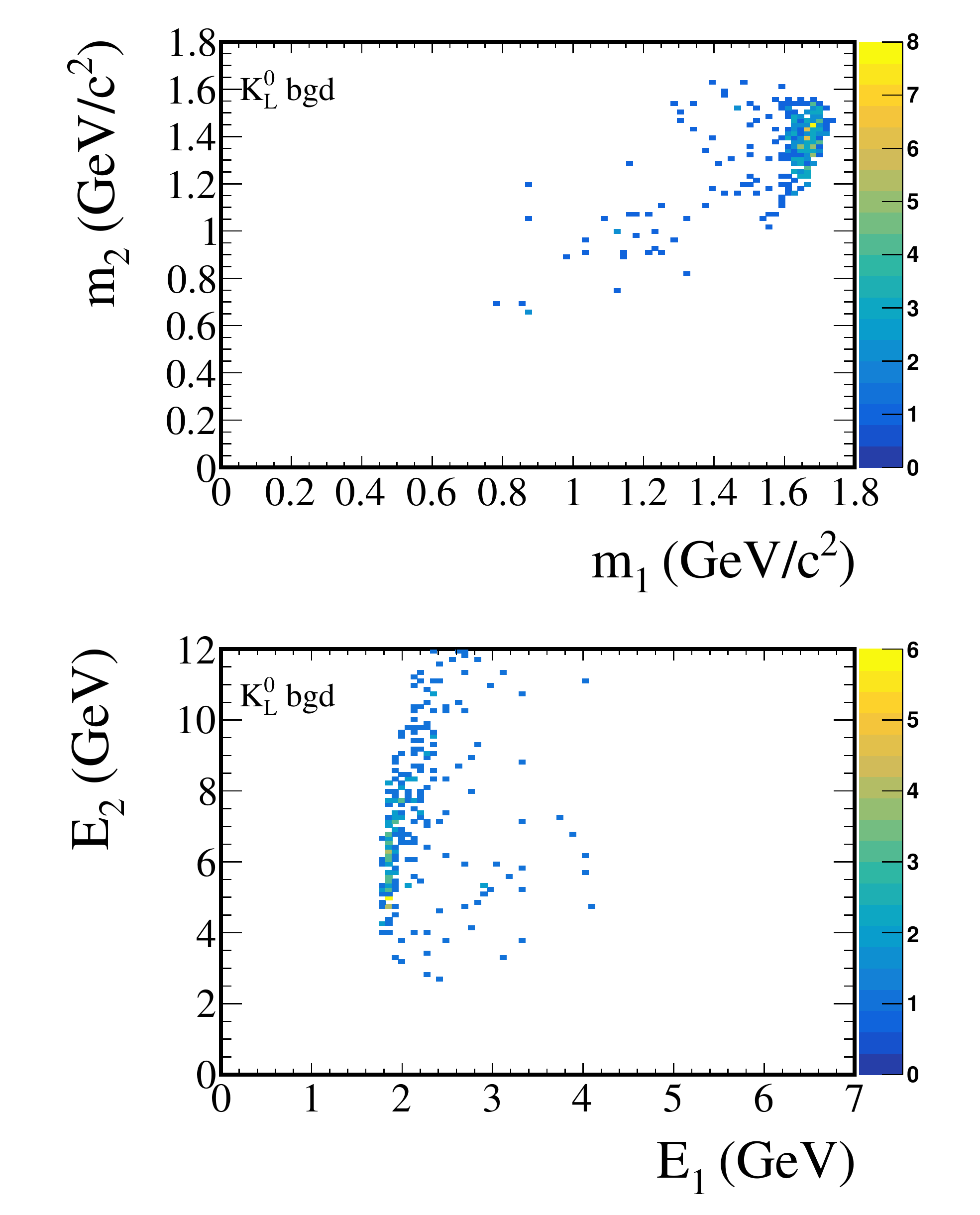}
    \end{tabular}
    \caption{The two solutions for the computed neutralino mass (top plots) and $\tau$ energy (bottom plots) for $\tau^-\to\pi^-\neu$, $\neu\to\eta\nu$, $\eta\to\pi^+\pi^-\pi^0$  signal events with $m_{\neu}=1.1$~GeV (left plots) and for  $\tau^-\to \pi^-\pi^0_p K_L \nu_\tau$ background events (right plots).}
    \label{fig:Eimi}
\end{figure}

As detailed in Ref.~\cite{Dib:2019tuj}, we further apply the constraints of the signal decay chain to determine the neutrino 4-momentum up to a 2-fold ambiguity. For each of the two solutions, indicated by $i=1,2$, this allows us to compute the calculated neutralino mass $m_i$ and the $\tau$ energy $E_i$ in the $e^+ e^-$ center-of-mass frame. 
The distributions of these variables are shown in Fig.~\ref{fig:Eimi} for signal events with $m_{\neu}=1.1$~GeV and for $\tau^-\to \pi^-\pi^0_p K_L \nu_\tau$ background events. The background plots contain 300 events, illustrating the full Belle~II dataset.
The $m_i$ distributions for signal, which peak at the generated mass, are missing the radiative tails seen in Ref.~\cite{Dib:2019tuj}, since our signal simulation lacks final-state radiation. The signal $E_i$ distributions, which peak at half the center-of-mass energy, show tails that result from the simulated initial-state radiation. The background distributions are very different from those of the signal: events accumulate at high values of $m_i$ and away from the correct values of $E_i$. Therefore, we conclude that the analysis is essentially background-free, with less background than even in the case of Ref.~\cite{Dib:2019tuj}. 
%%%%%%%%%

A related background channel in the $\neu\to \eta\nu_\tau$ signal channel is $\tau^-\to \pi^-\pi^0_p K_L \nu_\tau$ followed by $K_L\to \pi^\pm\ell^\mp\nu$. In this case, only the prompt $\pi^0$ is available to fake the $\eta$ signal, resulting in a factor of 3 suppression relative to the background from $K_L\to \pi^+\pi^-\pi^0$. This background is further suppressed to the sub-percent level by rejecting DVs formed by a lepton. Overall, its contribution is expected to be smaller than that of $K_L\to \pi^+\pi^-\pi^0$.

Background in the $\neu\to \eta'\nu_\tau$ channel is much smaller than in the $\neu\to \eta\nu_\tau$ channel. This arises from the higher mass of the $\eta'$, as well as the fact that the dominant decays $\eta'\to\eta\pi^+\pi^-$ and $\rho\gamma$ provide an additional intermediate hadron ($\eta$, and to some extent, $\rho$) for which a mass cut can be used for background rejection. 

In the $\lambda'_{312}$ case, there is no $\tau$-decay background that leads to the same final state as the $\neu\to K^{*0}\nu_\tau$ signal. Therefore, the background for this channel is much smaller than our estimate for $\neu\to \eta\nu_\tau$. The $\neu\to K_S\nu_\tau$ channel suffers from $\tau^-\to \pi^- K_S \nu_\tau$ background, which can be suppressed by requiring that the $K_S$ momentum vector be inconsistent with the direction between the $\tau^-$ and $K_S$ decay vertices. The neutralino flight length must be long enough for this cut to be effective. Therefore, in Sec.~\ref{sec:results} we present results with the $\neu\to K_S\nu_\tau$ channel only for $c\tau_{\neu}>10$~cm.

\section{Sensitivity estimation method}\label{sec:sensitivity}

In what follows we estimate the sensitivity of the Belle~II experiment for the $\tau \rightarrow  \neu M_1^{(*)}$, $\neu\to M_2^{(*)}\nu_\tau$ signal using $e^- e^+ \rightarrow \tau^+ \tau^-$ events. First, we consider the visible branching fraction of the neutralino, defined to be the total branching fraction into final states with at least two charged pions, which are necessary for finding the DV. 
In the $\lambda'_{311} \ne 0$ case, we have
\begin{eqnarray}
    \Br(\tilde{\chi}_1^0 \rightarrow \text{ visibles})  =  && \Br(\neu \rightarrow \nu_\tau \pi^0)\cdot \Br(\pi^0 \rightarrow \text{ visibles})  + \Br(\neu \rightarrow \nu_\tau \rho^0)\cdot \Br(\rho^0 \rightarrow \text{ visibles}) \nonumber \\
    && + \Br(\neu \rightarrow \nu_\tau \eta)\cdot \Br(\eta \rightarrow \text{ visibles}) + \Br(\neu \rightarrow \nu_\tau \eta')\cdot \Br(\eta' \rightarrow \text{ visibles}) \nonumber \\
    && + \Br(\neu \rightarrow \nu_\tau \omega)\cdot \Br(\omega \rightarrow \text{ visibles}),
\end{eqnarray}
where the visible decay branching ratios of the mesons are
\begin{eqnarray}
    \Br(\pi^0 \rightarrow \text{ visibles})  &=&  0,\nonumber\\
    \Br(\rho^0 \rightarrow \text{ visibles}) &=& 1, \nonumber\\
    \Br(\omega \rightarrow \text{ visibles}) &=& \Br(\omega\rightarrow \pi^+ \pi^-) + \Br(\omega\rightarrow \pi^+ \pi^- \pi^0) ,  \nonumber \\
    \Br(\eta \rightarrow \text{ visibles}) &=& \Br(\eta \rightarrow \pi^+ \pi^- \pi^0) + \Br(\eta \rightarrow \pi^+ \pi^- \gamma), \nonumber\\
    \Br(\eta' \rightarrow \text{ visibles}) &=& \Br(\eta' \rightarrow \pi^+ \pi^- \eta) + \Br(\eta' \rightarrow \rho^0 \gamma)\cdot \Br(\rho^0 \rightarrow \text{ visibles}) \\
    &&+ \Br(\eta' \rightarrow \pi^0 \pi^0 \eta)\cdot\Br(\eta \rightarrow \text{ visibles})+ \Br(\eta' \rightarrow \omega \gamma)\cdot\Br(\omega \rightarrow \text{ visibles}).\nonumber
\end{eqnarray}
%%%%%%%
In $\lambda'_{312} \ne 0$ scenario,
the `visible' decay branching ratio of the neutralino is
\begin{eqnarray}
    \Br(\tilde{\chi}_1^0 \rightarrow \text{ visibles})  = & \Br(\neu \rightarrow \nu_\tau K^{*0})\cdot \Br(K^{*0} \rightarrow \text{ visibles}) \nonumber\\
    &+ \Br(\neu \rightarrow \nu_\tau K_S)\cdot \Br(K_S \rightarrow \text{ visibles}) ,
\end{eqnarray}
with
\begin{eqnarray}
    \Br(K^{*0} \rightarrow \text{ visibles}) &=& \Br(K^{*0}\rightarrow K^\pm \pi^\mp),  \label{eq:312-n1-DV} \\
    \Br(K_S \rightarrow \text{ visibles}) &=& \Br(K_S\rightarrow \pi^+ \pi^- ). \label{eq:312-KS-DV}
\end{eqnarray}
A summary of the meson branching fractions is given in Table~\ref{tab:mesonbrs}.

Next, we consider the signal acceptance and efficiency. Given the similarity between the search proposed here and the one discussed in Ref.~\cite{Dib:2019tuj}, we use that reference to define the event-selection criteria and their related efficiencies. In principle, a DV can be reconstructed if its position is within an acceptance volume, which we take to be effectively defined by the radial and longitudinal requirements $10<r<80$~cm, $-40<z<120$~cm. The $r>10$ cm cut rejects most DVs that arise from material interaction and $K_S$ decays. The other requirements ensure a sufficient number of hits for adequate track and vertex reconstruction, given the size of the Belle~II tracking systems.

For a DV within the acceptance volume, the efficiency to reconstruct a signal event is 12\%, including the impact of reconstructing the two prompt tracks and application of particle-identification criteria on both tracks~\cite{Dib:2019tuj}. 
Relying on particle-reconstruction efficiencies at the $B$-factories~\cite{Allmendinger:2012ch,TheBABAR:2013jta}, we take the efficiency to find any additional $\pi^0$ or photon that is part of the signal decay to be 70\%, and the efficiency for an additional pair of charged pions to be 85\%.

While generally, the position of the DV is the decay position of the neutralino, this is not the case when $M_2=K_S$, because of the significant proper flight distance of the $K_S$, $c\tau_{K_S}=2.7$~cm. As a result, the $M_2=K_S$ mode is susceptible to background such as $\tau^-\to \pi^- K_S \nu_\tau$. This background will be suppressed by requiring that the $K_S$ momentum does not point back to the IP. The efficiency of this requirement is taken to be 90\%.

For each value of the neutralino mass $m_{\neu}$ and the relevant coupling $\lambda'_{31k}$, we calculate the total number of signal events observed in the experiment,
\begin{equation}
N_S = 2 N_{\tau^- \tau^+} \cdot \Br(\tau \rightarrow \text{ 1 prong}) \cdot \Br(\tau \rightarrow \neu M_1^{(*)})    \cdot \Br(\neu \rightarrow \text{ visibles}) \cdot \epsilon_{\text{acc.}} \cdot  \epsilon_{\text{det.}},
\end{equation}
where $\Br(\tau \rightarrow \text{ 1 prong})\approx 85\%$ is the branching fraction for $\tau$ decays into a single track, $M_1^{(*)}$ are listed in Table~\ref{tab:final-states-summary}, $\Br(\tau \rightarrow \neu M_1^{(*)})$ and $\Br(\neu \rightarrow \text{ visibles})$ are 
calculated as described above,
$\epsilon_{\text{acc.}}$ denotes the acceptance, and $\epsilon_{\text{det.}}$ is the reconstruction efficiency discussed above.

The acceptance $\epsilon_{\text{acc.}}$ depends on the neutralino boost, lifetime, and travel direction, as well as the geometry of the acceptance volume.
We estimate $\epsilon_{\text{acc.}}$ using a MC simulation with the Pythia 8.243~\cite{Sjostrand:2006za,Sjostrand:2007gs} event generator. We use the Pythia module \texttt{WeakSingleBoson:ffbar2ffbar(s:gm)} to produce $N_{\text{MC}}\sim 1.5\times 10^4$ $e^- e^+ \rightarrow \tau^- \tau^+$ events, including simulation of initial-state radiation.
All the simulated $\tau$ leptons undergo the signal decays listed in Table~\ref{tab:final-states-summary} exclusively, according to the relative branching ratios computed with the formulas given in Sec.~\ref{sec:prodANDdecay}.
We estimate $\epsilon_{\text{acc.}}$ with
\begin{eqnarray}
\epsilon_{\text{acc.}} &=& \frac{1}{2\,N_{\text{MC}}} \sum_{i=1}^{2\,N_{\text{MC}}} \epsilon_{\text{acc.}}^{i},
\end{eqnarray}
where $\epsilon_{\text{acc.}}^{i}$ is the acceptance for the polar angle $\theta_i$ of the $i^{\text{th}}$ simulated neutralino, obtained from
\begin{eqnarray}
\epsilon_{\text{acc.}}^{i} &=& e^{-z^I_i/\lambda_i^z} \cdot (1  -   e^{-z^O_i/\lambda_i^z}).
\end{eqnarray}
Here $z^I_i$ is the $z$ coordinate at which the neutralino enters the acceptance volume, and 
$z^O_i$ is the distance traveled inside the acceptance volume:
\begin{eqnarray}
z^I_i                          &\equiv& \text{min}(Z, |R_I/\tan{\theta_i}|),\\
z^O_i                         &\equiv& \text{min}(Z,|R_O/\tan{\theta_i}|) - z^I_i,
\end{eqnarray}
where $R_I=10$~cm, $R_O=80$~cm are the inner and outer radii of the acceptance volume, $Z=120$~cm for $\tan\theta_i>0$ and $40$~cm for $\tan\theta_i<0$ are the longitudinal edges of the acceptance volume.
The factor
\begin{eqnarray}
\lambda_i^z \equiv \beta_i^z \, \gamma_i \, c \, \tau_{\neu} ,
\end{eqnarray}
is the average flight distance in the $z$ direction of the neutralino given the lifetime $\tau_{\neu}$ and the $z$-direction boost factor $\beta_i^z \, \gamma_i$ obtained from the Pythia simulation.

We evaluate $N_S$ in a rectangular grid of points in the parameter space $m_{\neu}$ vs. $\lambda'_{31k}/m^2_{\tilde{f}}$ for $k=1,2$. 
In the $k=1$ case, we scan $m_{\neu}$ in steps of 0.01 GeV for $m_{\neu}>0.55$~GeV, while for $0.547<m_{\neu}<0.55$~GeV
we take the finer steps of 0.001~GeV.
In the $k=2$ case, we scan $m_{\neu}$ in steps of 0.01~GeV for $0.5 <m_{\neu}< 1.28$~GeV, and in steps of 0.001~GeV in the ranges  $0.49 <m_{\neu}< 0.499$~GeV and $1.28  <m_{\neu}< 1.3$~GeV.
The ratios $\lambda'_{31k}/m^2_{\tilde{f}}$ are scanned between $10^{-10}$ GeV$^{-2}$ and $8\times 10^{-4}$ GeV$^{-2}$, for 49 points in total.

\begin{figure}[!htbp]
\centering
\includegraphics[width=0.49\textwidth]{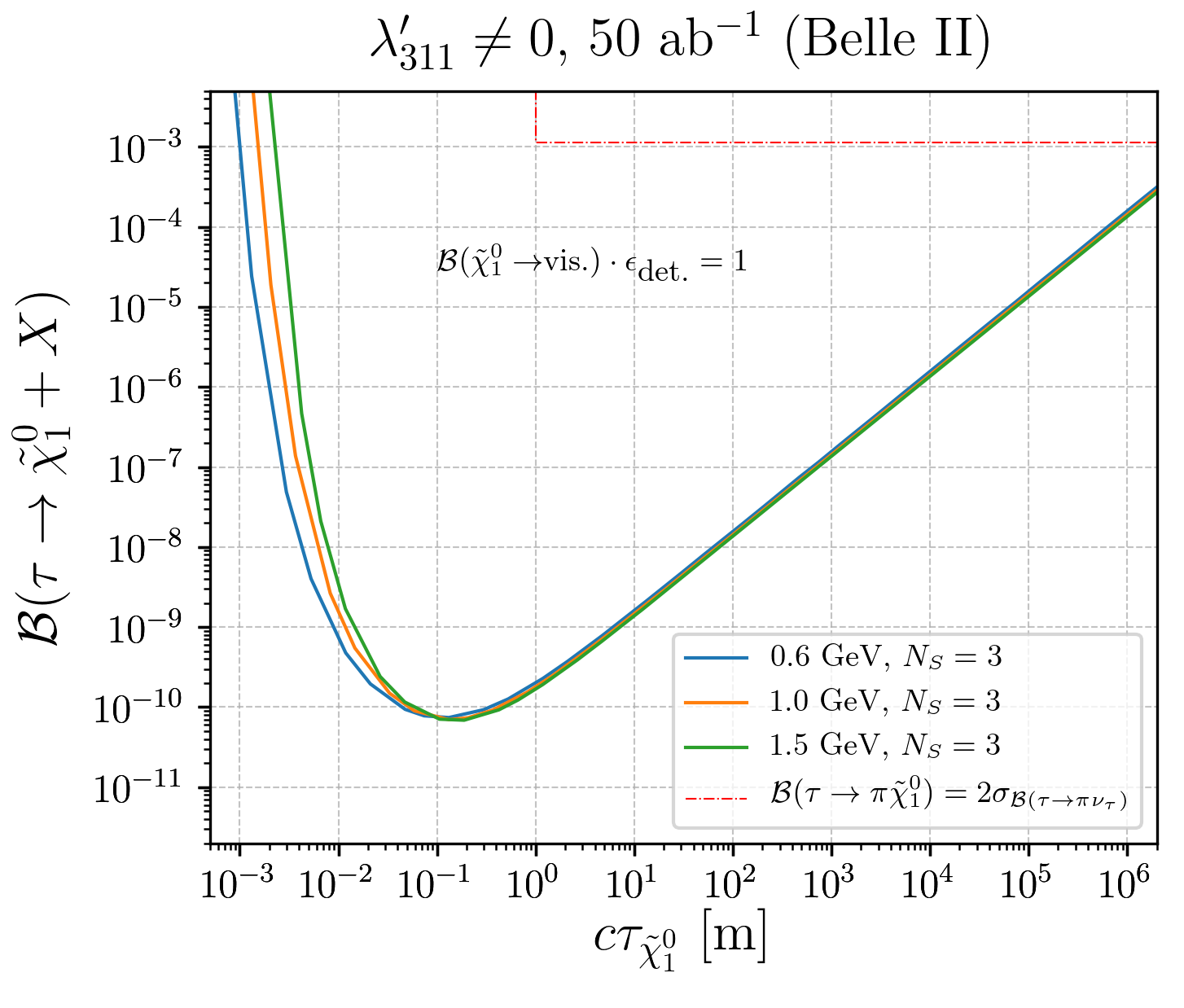}
\includegraphics[width=0.49\textwidth]{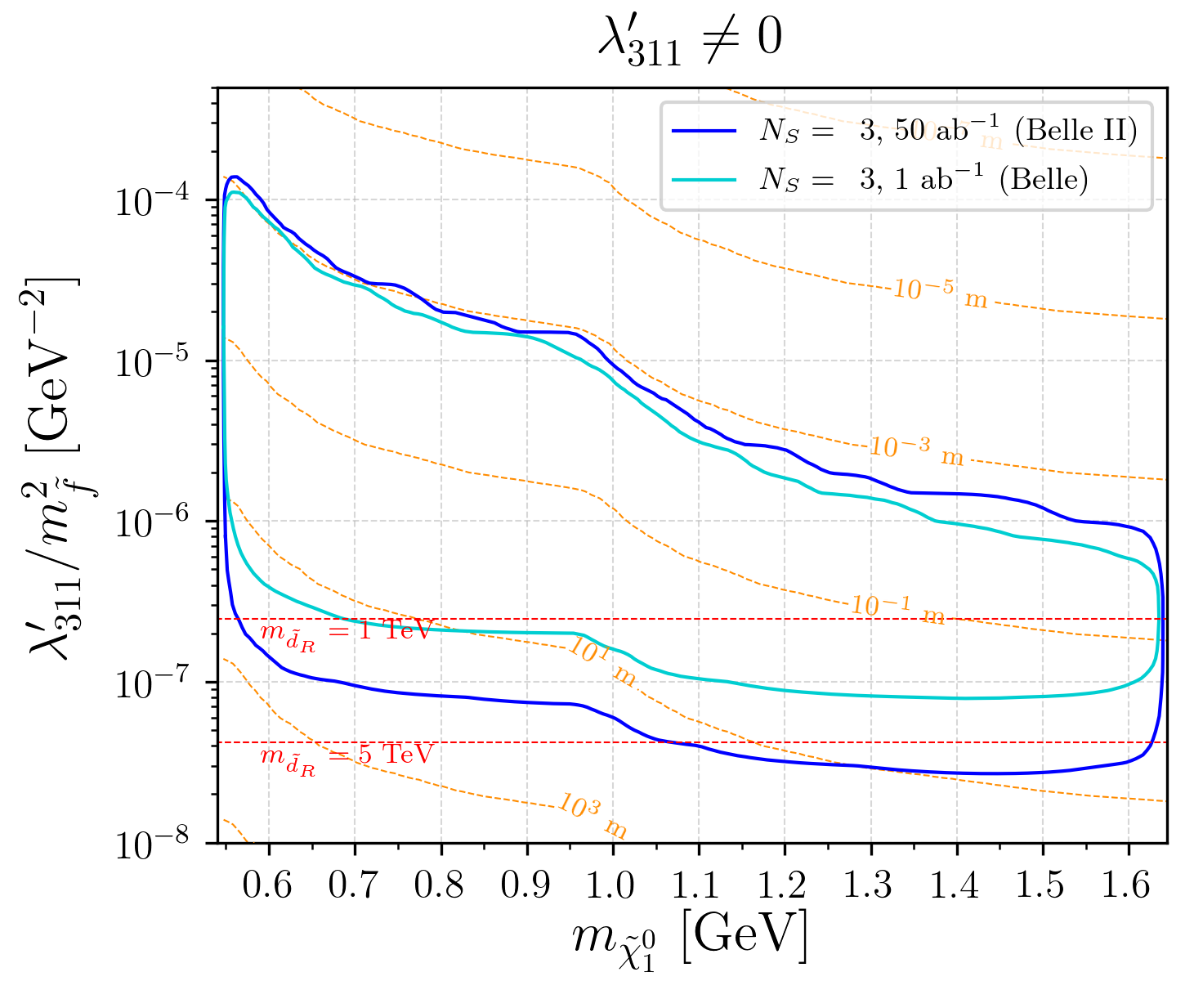}
\caption{Sensitivity reach for the case  $\lambda'_{311}\ne 0$. Curves show the parameter values that yield observation of 3 signal events, corresponding to 95\% confidence-level limits for negligible background.
\textit{Left}: Model-independent limits on the branching fraction $\Br(\tau \to M_1^{(*)}\neu )$, taking 100\% for the branching fraction  $\neu\to\text{visibles}$ and the detector efficiency $\epsilon_{\text{det.}}$, as a function of $c\tau_{\neu}$.
Three curves correspond to $m_{\neu}= 0.6$ (blue), 1.0 (orange), and 1.5~GeV (green). 
The red dot-dashed line represents the current constraint that arises from the experimental and theoretical uncertainties in  $\Br(\tau^- \rightarrow \pi^- \nu_\tau)$, Eq.~(\ref{eq:tauBRlimit}). This constraint is relevant only for large lifetimes, and is drawn only for $c\tau_{\neu}>1$~m.
\textit{Right}: model-dependent results on $\lambda'_{311}/m^2_{\tilde{f}}$ vs. $m_{\neu}$ shown for the Belle (light blue, $1~\text{ab}^{-1}$) and Belle~II (dark blue, $50~\text{ab}^{-1}$) data sets.
The orange dashed curves show the proper decay length, $c\tau_{\neu}$, of the neutralino in intervals of \mbox{100 m}.
The red dashed horizontal lines denote the current limits on  $\lambda'_{31k}/m^2_{\tilde{q}_R}$ (Eq.~\eqref{eq:rpvlimit31k}) for squark masses of 1~TeV and 5~TeV.
}
\label{fig:311_sensitivities}
\end{figure}

\section{Numerical results}
\label{sec:results}
We proceed to discuss the numerical results for both model-independent and model-dependent limits. 
The model-independent limits are quoted in terms of the branching fraction $\Br(\tau \to \neu M_1^{(*)})$, taking the branching fraction $\Br(\neu\to \text{visibles})$ and detector efficiency $\epsilon_{\text{det.}}$ to be 100\%. These limits are shown as a function of $c\tau_{\neu}$, requiring only that the $\neu$ decay within the acceptance region. 
The model-dependent limits are given in terms of the ratio between the RPV coupling and the universal squark mass squared, $\lambda'_{31k}/m^2_{\tilde{f}}$, as a function of $m_{\neu}$, and use the efficiency estimates described above. 
Both types of limits are obtained by requiring observation of $N_S=3$ signal events at Belle~II with a data sample of $50~\text{ab}^{-1}$ for a background expectation of close to zero events, corresponding to 95\% confidence-level limits. The model-dependent limits are presented also for an integrated luminosity of $1~\text{ab}^{-1}$, corresponding to the data already collected by the Belle experiment, and twice the integrated luminosity of BABAR.
We note that for small values of  $\lambda'_{311}/m^2_{\tilde{f}}$, the Belle limits are weaker than the Belle~II limits by about the $4^\text{th}$ root of the ratio of their integrated luminosities, $50^{1/4}\sim 2.6$. This is because in the large decay length limit, the number of signal events is proportional to $(\lambda'_{311}/m^2_{\tilde{f}})^4$.

In Fig.~\ref{fig:311_sensitivities} we show the limits for $\lambda'_{311}\ne 0$. 
The model-independent limits are given in the left panel for three neutralino mass values: 0.6 GeV (blue), 1.0  GeV (orange), and 1.5 GeV (green) using the full Belle~II data.
Also plotted is the limit extracted from the current uncertainties in the measured branching fraction for $\tau^-\to \pi^-\nu_\tau$ and its theoretical prediction (red dot-dashed lines), Eq.~(\ref{eq:tauBRlimit}).
We find that Belle~II may probe $\Br(\tau \to \neu + X)$ values many orders of magnitude smaller than the present experimental upper bound, for $m_{\neu}$ below $m_{\tau}$, with the best sensitivity obtained for  $c\tau_{\neu}\sim 10-20$ cm.

The right panel of Fig.~\ref{fig:311_sensitivities} displays the model-dependent limits on $\lambda'_{31k}/m^2_{\tilde{f}}$ that we expect with the full Belle and Belle~II data sets.
Also shown is the upper bound of Eq.~\eqref{eq:rpvlimit31k} on the ratio $\lambda'_{311}/m^2_{\tilde{q}_R}$ for the benchmark squark masses of 1~and 5~TeV. 
Since the limit of Eq.~\eqref{eq:rpvlimit31k} has a different power of $m_{\tilde{f}}$, its comparison to our results depends on the assumed value for $m_{\tilde{f}}$.
The upper bound derived from the uncertainty on $\tau^-\to\pi^-\neu$ is irrelevant here, as it corresponds to values of $\lambda'_{311}/m_{\tilde f}^2$ and $m_{\neu}$ where $c \tau_{\neu}< 1\mbox{ m}$, in which this bound does not apply.
We find that the search we propose is more sensitive than the limit of Eq.~\eqref{eq:rpvlimit31k} for sfermion masses of up to about 5~TeV.
The right plot of Fig.~\ref{fig:311_sensitivities} also shows orange dashed isocurves of the proper decay length $c\tau_{\neu}$ of the neutralino in the shown parameter space.

\begin{figure}[!htbp]
\centering
\includegraphics[width=0.49\textwidth]{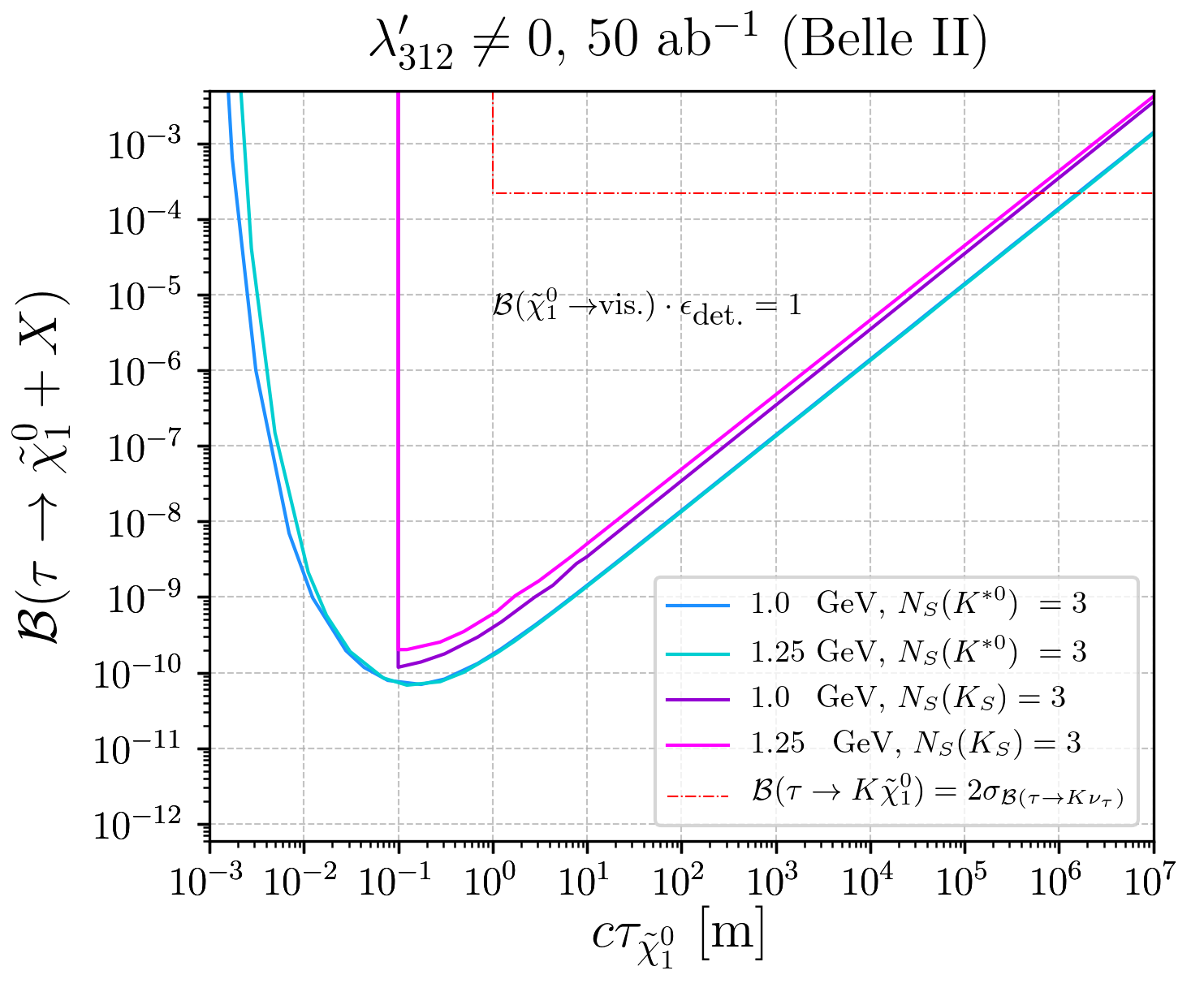}
\includegraphics[width=0.49\textwidth]{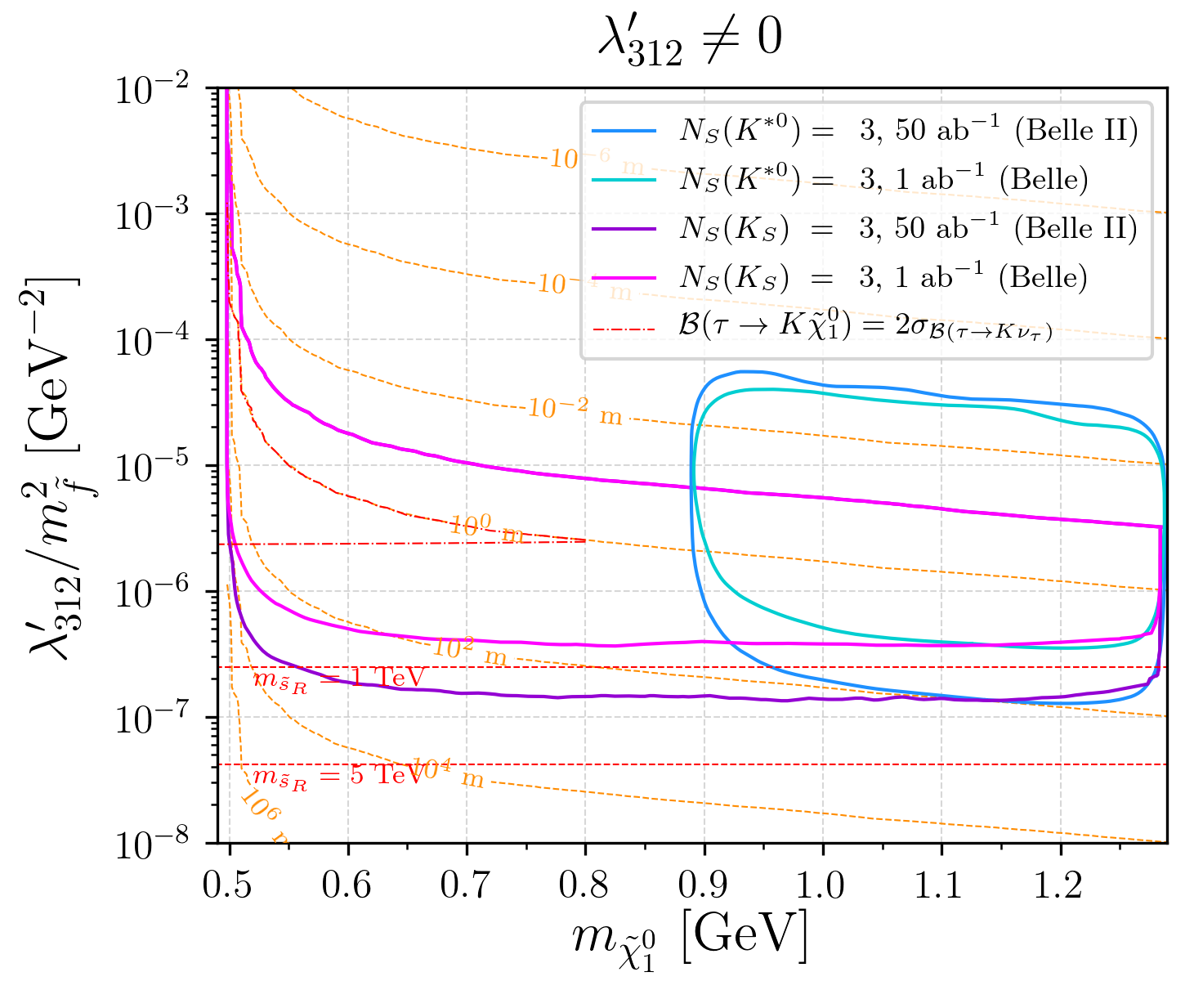}
\caption{Sensitivity reach for the case $\lambda'_{312} \ne 0$. See caption of Fig.~\ref{fig:311_sensitivities} for details. Unlike in Fig.~\ref{fig:311_sensitivities}, the red dot-dashed curve that represents the constraint from  $\Br(\tau^- \rightarrow \pi^- \nu_\tau)$ (Eq.~(\ref{eq:tauBRlimit})) is relevant also for the model-dependent limits shown in the right panel for $c\tau_{\neu}>1$~m. 
The curves for $\neu\to K_S \nu_\tau$ are shown only for $c\tau_{\neu}>10$~cm, because of the high background at short flight distances.
}
\label{fig:312_sensitivities}
\end{figure}

Figure~\ref{fig:312_sensitivities} gives the results for the $\lambda'_{312}\ne 0$ scenario, treating separately the $M_2=K_S$ and $M_2^*=K^{*0}$ cases.
We overlap the limits derived from $\tau \to K \nu_\tau$ for $c\tau\gtrsim 1$ m only. As mentioned above, the $K_S$ channel suffers from large background at short distances. We incorporate this by showing the limits expected for the $K_S$ channel only for $c\tau_{\neu}>10$~cm (magenta and purple curves in the two panels). Despite this shortcoming of the $K_S$ channel, its advantage is that it is sensitive to $m_{\neu}<0.9$~GeV. 
Overall, comparison of Figs.~\ref{fig:311_sensitivities} and~\ref{fig:312_sensitivities} shows that the scenario $\lambda'_{311}\ne 0$ is probed with greater sensitivity, by almost an order of magnitude, than the scenario $\lambda'_{312}\ne 0$.

\section{Conclusions}\label{sec:conclu}

In recent years, searches for long-lived particles (LLPs) have become an important means for probing new physics, particularly at the LHC.
A well-motivated scenario for new physics is R-parity-violating (RPV) supersymmetry with light, long-lived neutralinos.
Such neutralinos may be produced at colliders and lead to an exotic displaced-vertex signature.
In this work, we develop a method for studying the scenario in which the neutralino mass $m_{\neu}$ is roughly between 0.5 GeV and $m_{\tau}-m_\pi$.
We study two benchmark scenarios in which one of the RPV operators, $\lambda'_{311} \ L_3 Q_1 \bar{D}_1$ or $\lambda'_{312} \ L_3 Q_1 \bar{D}_2$, is non-vanishing.
Each of these operators induces $\tau$ decay into the neutralino plus an accompanying meson $M_1^{(*)}$, and also the neutralino decay into a meson $M_2^{(*)}$ plus a tau neutrino.
We propose to conduct the search at electron-positron $B$-factories, which are particularly suitable for the study of rare $\tau$ decays. Specifically, the BABAR and Belle experiments have collected a combined data sample of $1.4\times 10^9$ $e^+ e^-\to \tau^+ \tau^-$ events, and the ongoing Belle~II experiment is scheduled to collect $4.6\times 10^{10}$ events in the coming years.

For both RPV operators, we consider in detail the possible decays of the $\tau$ lepton and of the neutralino, defining the final states and displaced-vertex signatures that we propose to study.
Correspondingly, we perform Monte-Carlo simulations to determine the leading background and the signal events acceptance. In particular, simulation is also used to estimate the largest source of background, $\tau^-\to\pi^-K_L \nu_\tau$ with $K_L\to \pi^+\pi^-\pi^0$, which impacts the $\neu\to\eta\nu_\tau$, $\eta\to \pi^+\pi^-\pi^0$ signal channel. After using the constraints of the signal decay to obtain the neutralino mass and the $\tau$ energy, we expect the effective background yield in this channel to be negligible even in the full Belle~II sample. In the other major channels, the expected background is even smaller.

Using our estimated acceptance, efficiency, and estimated negligible background yield, we present numerical results for both model-independent and model-dependent limits.
In both scenarios, the model-independent limits on $\Br(\tau \rightarrow X \neu)$ are orders of magnitude tighter than the current bounds stemming from the combined experimental and theoretical uncertainties in the branching ratio for $\tau$ decay into a charged hadron plus a neutrino.
The model-dependent limits on $\lambda'_{311}/m^2_{\tilde{f}}$ are up to two orders of magnitude tighter than the limits obtained by recasting~\cite{Bansal:2019zak} the $pp\to W' \to \tau\nu_\tau$ search for squark mass $m_{\tilde{q}_R}=1$~TeV, and comparable for $m_{\tilde{q}_R}=5$~TeV.
The limits on $\lambda'_{312}/m^2_{\tilde{f}}$ are comparable with the $pp\to W' \to \tau\nu_\tau$ for $m_{\tilde{q}_R}=1$~TeV. We note that the bounds obtained from the $pp\to W' \to \tau\nu_\tau$ search are relevant only when the RPV process involves a virtual squark, while our results also probe the case of a virtual stau and sneutrino.

Together with Ref.~\cite{Dib:2019tuj}, this work exemplifies the usefulness of $B$-factory experiments for searching for new physics related to the $\tau$ lepton, e.g., long-lived particles produced in $\tau$ decays.
We propose that further scenarios be pursued along this direction.

\begin{appendices}

\section{Meson decay constants and branching fractions}\label{app:constants}

The decay constant for the pseudoscalar current is \cite{deVries:2015mfw}
\begin{equation}
f_M^S \equiv
\left<0|\bar q_1 \gamma^5 q_2 | M \right> = i\frac{m_{M}^2}{m_{q_1}+m_{q_2}} f_M,\label{eq:decay_const_scalar}
\end{equation}
where $M$ is the meson mass and $q_1$ and $q_2$ its valence quarks.
The values of $f_M$ used in Sec.~\ref{sec:prodANDdecay} are summarized in Table \ref{tab:decayconst}.
Because the tensor decay constants $f_{M^*}^T$ are not known for all mesons, following Ref.~\cite{deVries:2015mfw}, we assume $f_{M^*}^T\approx f_M$.
Note that for the neutral $\pi$ and $\rho$ mesons, $f_{\pi^0,\rho^0} = f_{\pi,\rho}/\sqrt{2}$ in Eq.~\ref{eq:decay_const_scalar}.
For the $\eta$ and $\eta'$ mesons, the decay constants are obtained from Ref.~\cite{0905.2051}, which takes into account the mixing between the $\eta^0$ and $\eta^8$ SU(3) flavor states.
Distinct calculations of the $a_1(1260)$ meson decay constant find different values around the $\rho$ decay constant \cite{Bloch:1999vka,Cheng:2003bn}.
For definiteness, we chose $f_{a_1}=f_\rho$.
\begin{table}[htb]
\centering
\begin{tabular}{lcc}
\hline
Constant & Value [MeV] & Ref.\\
\hline\hline
$f_{\pi^\pm}$ & 130.2 & \cite{Zyla:2020zbs}\\
$f_\rho$ & 209 & \cite{Ebert:2006hj}\\
$f_{a_1}$ & 209 & \cite{Bloch:1999vka,Cheng:2003bn}\\
$f_{K^\pm}$ & 155.7 & \cite{Zyla:2020zbs}\\
$f_{K^*}$ & 230 & \cite{Dreiner:2006gu} \\
$f_\omega$ & 195.3 &\cite{Ebert:2006hj}\\
$f_\eta^{\bar d d}$ & 49.42 & \cite{0905.2051}\\
$f_{\eta'}^{\bar d d}$ & 111.43 & \cite{0905.2051}\\
\hline
\hline
\end{tabular}
\caption{Values of the decay constants used in this work.}\label{tab:decayconst}
\end{table}

A summary of the relevant meson branching fractions is given in Table~\ref{tab:mesonbrs}.

\begin{table}[tb]
    \centering
    \begin{tabular}{c|c}
    Decay & Branching fraction\\
    \hline\hline
 $\omega \to \pi^+\pi^-\pi^0$ & 0.893\\
$\omega \to \pi^+\pi^-$      & 0.0153\\
$\eta \to \pi^+\pi^-\pi^0$   & 0.2292\\
$\eta \to \pi^+\pi^-\gamma$  & 0.0422\\
$\eta \to \gamma \gamma$     & 0.3941\\
$\eta \to \pi^0\pi^0\pi^0$   & 0.3268\\
$\eta'\to \pi^+\pi^-\eta$    & 0.426\\
$\eta'\to \rho\gamma$        & 0.289\\
$\eta'\to \pi^0\pi^0\eta$    & 0.228\\
$\eta'\to \omega\gamma$      & 0.0262\\
        \hline
        $K_S\to \pi^+\pi^-$ & 0.692\\
        $K^{*0}\to K^\pm \pi^\mp$ & 0.665\\ %0.6649467
        \hline\hline
    \end{tabular}
    \caption{Branching fractions for the meson decays considered in our analysis. The values are reproduced from Ref.~\cite{Zyla:2020zbs}.}
    \label{tab:mesonbrs}
\end{table}

\end{appendices}

\section*{Acknowledgements:} 
We would like to thank Jordy de Vries, Marcela Gonz\'alez, Sergey Kovalenko, and Torbjörn Sjöstrand for useful discussions.
Z. S. W. is supported partly by the Ministry of Science and Technology (MoST) of Taiwan with grant number MoST-109-2811-M-007-509, and partly by the Ministry of Science, ICT \& Future Planning of Korea, the Pohang City Government, and the Gyeongsangbuk-do Provincial Government through the Young Scientist Training Asia Pacific Economic Cooperation program of the Asia Pacific Center for Theoretical Physics.
C. O. D. acknowledges support from FONDECYT (Chile) Grant No. 1170171 and ANID (Chile) PIA/APOYO AFB 180002.
J. C. H.  acknowledges support from grant FONDECYT (Chile) No.1201673.
N. A. N. was supported by ANID (Chile) under the grant ANID REC Convocatoria Nacional Subvenci\'on a Instalaci\'on en la Academia Convocatoria A\~no 2020, PAI77200092.

\bibliographystyle{JHEP}
\bibliography{main}

\providecommand{\href}[2]{#2}\begingroup\raggedright\begin{thebibliography}{10}

\bibitem{Nilles:1983ge}
H.~P. Nilles, {\it {Supersymmetry, Supergravity and Particle Physics}},  {\em
  Phys. Rept.} {\bf 110} (1984) 1--162.

\bibitem{Martin:1997ns}
S.~P. Martin, {\it {A Supersymmetry primer}},  {\em Adv. Ser. Direct. High
  Energy Phys.} {\bf 21} (2010) 1--153,
  [\href{http://arxiv.org/abs/hep-ph/9709356}{{\tt hep-ph/9709356}}].

\bibitem{Gildener:1976ai}
E.~Gildener, {\it {Gauge Symmetry Hierarchies}},  {\em Phys. Rev. D} {\bf 14}
  (1976) 1667.

\bibitem{Veltman:1980mj}
M.~Veltman, {\it {The Infrared - Ultraviolet Connection}},  {\em Acta Phys.
  Polon. B} {\bf 12} (1981) 437.

\bibitem{Dreiner:1997uz}
H.~K. Dreiner, {\it {An Introduction to explicit R-parity violation}},  {\em
  Adv. Ser. Direct. High Energy Phys.} {\bf 21} (2010) 565--583,
  [\href{http://arxiv.org/abs/hep-ph/9707435}{{\tt hep-ph/9707435}}].

\bibitem{Barbier:2004ez}
R.~Barbier et~al., {\it {R-parity violating supersymmetry}},  {\em Phys. Rept.}
  {\bf 420} (2005) 1--202, [\href{http://arxiv.org/abs/hep-ph/0406039}{{\tt
  hep-ph/0406039}}].

\bibitem{Mohapatra:2015fua}
R.~N. Mohapatra, {\it {Supersymmetry and R-parity: an Overview}},  {\em Phys.
  Scripta} {\bf 90} (2015) 088004, [\href{http://arxiv.org/abs/1503.06478}{{\tt
  arXiv:1503.06478}}].

\bibitem{Ibanez:1991pr}
L.~E. Ibanez and G.~G. Ross, {\it {Discrete gauge symmetries and the origin of
  baryon and lepton number conservation in supersymmetric versions of the
  standard model}},  {\em Nucl. Phys. B} {\bf 368} (1992) 3--37.

\bibitem{Dreiner:2012ae}
H.~K. Dreiner, M.~Hanussek, and C.~Luhn, {\it {What is the discrete gauge
  symmetry of the R-parity violating MSSM?}},  {\em Phys. Rev. D} {\bf 86}
  (2012) 055012, [\href{http://arxiv.org/abs/1206.6305}{{\tt
  arXiv:1206.6305}}].

\bibitem{Aaboud:2018doq}
{\bf ATLAS} Collaboration, M.~Aaboud et~al., {\it {Search for photonic
  signatures of gauge-mediated supersymmetry in 13 TeV $pp$ collisions with the
  ATLAS detector}},  {\em Phys. Rev. D} {\bf 97} (2018), no.~9 092006,
  [\href{http://arxiv.org/abs/1802.03158}{{\tt arXiv:1802.03158}}].

\bibitem{Sirunyan:2017nyt}
{\bf CMS} Collaboration, A.~M. Sirunyan et~al., {\it {Search for gauge-mediated
  supersymmetry in events with at least one photon and missing transverse
  momentum in pp collisions at $\sqrt{s} = $ 13 TeV}},  {\em Phys. Lett. B}
  {\bf 780} (2018) 118--143, [\href{http://arxiv.org/abs/1711.08008}{{\tt
  arXiv:1711.08008}}].

\bibitem{Sirunyan:2019mbp}
{\bf CMS} Collaboration, A.~M. Sirunyan et~al., {\it {Search for supersymmetry
  in final states with photons and missing transverse momentum in proton-proton
  collisions at 13 TeV}},  {\em JHEP} {\bf 06} (2019) 143,
  [\href{http://arxiv.org/abs/1903.07070}{{\tt arXiv:1903.07070}}].

\bibitem{Sirunyan:2019ctn}
{\bf CMS} Collaboration, A.~M. Sirunyan et~al., {\it {Search for supersymmetry
  in proton-proton collisions at 13 TeV in final states with jets and missing
  transverse momentum}},  {\em JHEP} {\bf 10} (2019) 244,
  [\href{http://arxiv.org/abs/1908.04722}{{\tt arXiv:1908.04722}}].

\bibitem{Aad:2020nyj}
{\bf ATLAS} Collaboration, G.~Aad et~al., {\it {Search for new phenomena in
  final states with large jet multiplicities and missing transverse momentum
  using $ \sqrt{s} $ = 13 TeV proton-proton collisions recorded by ATLAS in Run
  2 of the LHC}},  {\em JHEP} {\bf 10} (2020) 062,
  [\href{http://arxiv.org/abs/2008.06032}{{\tt arXiv:2008.06032}}].

\bibitem{Alimena:2019zri}
J.~Alimena et~al., {\it {Searching for long-lived particles beyond the Standard
  Model at the Large Hadron Collider}},  {\em J. Phys. G} {\bf 47} (2020),
  no.~9 090501, [\href{http://arxiv.org/abs/1903.04497}{{\tt
  arXiv:1903.04497}}].

\bibitem{Lee:2018pag}
L.~Lee, C.~Ohm, A.~Soffer, and T.-T. Yu, {\it {Collider Searches for Long-Lived
  Particles Beyond the Standard Model}},  {\em Prog. Part. Nucl. Phys.} {\bf
  106} (2019) 210--255, [\href{http://arxiv.org/abs/1810.12602}{{\tt
  arXiv:1810.12602}}].

\bibitem{Curtin:2018mvb}
D.~Curtin et~al., {\it {Long-Lived Particles at the Energy Frontier: The
  MATHUSLA Physics Case}},  {\em Rept. Prog. Phys.} {\bf 82} (2019), no.~11
  116201, [\href{http://arxiv.org/abs/1806.07396}{{\tt arXiv:1806.07396}}].

\bibitem{Choudhury:1995pj}
D.~Choudhury and S.~Sarkar, {\it {A Supersymmetric resolution of the KARMEN
  anomaly}},  {\em Phys. Lett. B} {\bf 374} (1996) 87--92,
  [\href{http://arxiv.org/abs/hep-ph/9511357}{{\tt hep-ph/9511357}}].

\bibitem{Choudhury:1999tn}
D.~Choudhury, H.~K. Dreiner, P.~Richardson, and S.~Sarkar, {\it {A
  Supersymmetric solution to the KARMEN time anomaly}},  {\em Phys. Rev. D}
  {\bf 61} (2000) 095009, [\href{http://arxiv.org/abs/hep-ph/9911365}{{\tt
  hep-ph/9911365}}].

\bibitem{Belanger:2002nr}
G.~Belanger, F.~Boudjema, A.~Pukhov, and S.~Rosier-Lees, {\it {A Lower limit on
  the neutralino mass in the MSSM with nonuniversal gaugino masses}},  in {\em
  {10th International Conference on Supersymmetry and Unification of
  Fundamental Interactions (SUSY02)}}, pp.~919--924, 12, 2002.
\newblock \href{http://arxiv.org/abs/hep-ph/0212227}{{\tt hep-ph/0212227}}.

\bibitem{Bottino:2002ry}
A.~Bottino, N.~Fornengo, and S.~Scopel, {\it {Light relic neutralinos}},  {\em
  Phys. Rev. D} {\bf 67} (2003) 063519,
  [\href{http://arxiv.org/abs/hep-ph/0212379}{{\tt hep-ph/0212379}}].

\bibitem{Belanger:2003wb}
G.~Belanger, F.~Boudjema, A.~Cottrant, A.~Pukhov, and S.~Rosier-Lees, {\it
  {Lower limit on the neutralino mass in the general MSSM}},  {\em JHEP} {\bf
  03} (2004) 012, [\href{http://arxiv.org/abs/hep-ph/0310037}{{\tt
  hep-ph/0310037}}].

\bibitem{Vasquez:2010ru}
D.~Albornoz~Vasquez, G.~Belanger, C.~Boehm, A.~Pukhov, and J.~Silk, {\it {Can
  neutralinos in the MSSM and NMSSM scenarios still be light?}},  {\em Phys.
  Rev. D} {\bf 82} (2010) 115027, [\href{http://arxiv.org/abs/1009.4380}{{\tt
  arXiv:1009.4380}}].

\bibitem{Calibbi:2013poa}
L.~Calibbi, J.~M. Lindert, T.~Ota, and Y.~Takanishi, {\it {Cornering light
  Neutralino Dark Matter at the LHC}},  {\em JHEP} {\bf 10} (2013) 132,
  [\href{http://arxiv.org/abs/1307.4119}{{\tt arXiv:1307.4119}}].

\bibitem{Gogoladze:2002xp}
I.~Gogoladze, J.~D. Lykken, C.~Macesanu, and S.~Nandi, {\it {Implications of a
  Massless Neutralino for Neutrino Physics}},  {\em Phys. Rev. D} {\bf 68}
  (2003) 073004, [\href{http://arxiv.org/abs/hep-ph/0211391}{{\tt
  hep-ph/0211391}}].

\bibitem{Dreiner:2009ic}
H.~K. Dreiner, S.~Heinemeyer, O.~Kittel, U.~Langenfeld, A.~M. Weber, and
  G.~Weiglein, {\it {Mass Bounds on a Very Light Neutralino}},  {\em Eur. Phys.
  J. C} {\bf 62} (2009) 547--572, [\href{http://arxiv.org/abs/0901.3485}{{\tt
  arXiv:0901.3485}}].

\bibitem{Grifols:1988fw}
J.~Grifols, E.~Masso, and S.~Peris, {\it {Photinos From Gravitational
  Collapse}},  {\em Phys. Lett. B} {\bf 220} (1989) 591--596.

\bibitem{Ellis:1988aa}
J.~R. Ellis, K.~A. Olive, S.~Sarkar, and D.~Sciama, {\it {Low Mass Photinos and
  Supernova \{SN1987A\}}},  {\em Phys. Lett. B} {\bf 215} (1988) 404--410.

\bibitem{Lau:1993vf}
K.~Lau, {\it {Constraints on supersymmetry from SN1987A}},  {\em Phys. Rev. D}
  {\bf 47} (1993) 1087--1092.

\bibitem{Dreiner:2003wh}
H.~Dreiner, C.~Hanhart, U.~Langenfeld, and D.~R. Phillips, {\it {Supernovae and
  light neutralinos: SN1987A bounds on supersymmetry revisited}},  {\em Phys.
  Rev. D} {\bf 68} (2003) 055004,
  [\href{http://arxiv.org/abs/hep-ph/0304289}{{\tt hep-ph/0304289}}].

\bibitem{Dreiner:2013tja}
H.~K. Dreiner, J.-F. Fortin, J.~Isern, and L.~Ubaldi, {\it {White Dwarfs
  constrain Dark Forces}},  {\em Phys. Rev. D} {\bf 88} (2013) 043517,
  [\href{http://arxiv.org/abs/1303.7232}{{\tt arXiv:1303.7232}}].

\bibitem{Profumo:2008yg}
S.~Profumo, {\it {Hunting the lightest lightest neutralinos}},  {\em Phys. Rev.
  D} {\bf 78} (2008) 023507, [\href{http://arxiv.org/abs/0806.2150}{{\tt
  arXiv:0806.2150}}].

\bibitem{Dreiner:2011fp}
H.~K. Dreiner, M.~Hanussek, J.~S. Kim, and S.~Sarkar, {\it {Gravitino cosmology
  with a very light neutralino}},  {\em Phys. Rev. D} {\bf 85} (2012) 065027,
  [\href{http://arxiv.org/abs/1111.5715}{{\tt arXiv:1111.5715}}].

\bibitem{Hooper:2002nq}
D.~Hooper and T.~Plehn, {\it {Supersymmetric dark matter: How light can the LSP
  be?}},  {\em Phys. Lett. B} {\bf 562} (2003) 18--27,
  [\href{http://arxiv.org/abs/hep-ph/0212226}{{\tt hep-ph/0212226}}].

\bibitem{Bottino:2011xv}
A.~Bottino, N.~Fornengo, and S.~Scopel, {\it {Phenomenology of light
  neutralinos in view of recent results at the CERN Large Hadron Collider}},
  {\em Phys. Rev. D} {\bf 85} (2012) 095013,
  [\href{http://arxiv.org/abs/1112.5666}{{\tt arXiv:1112.5666}}].

\bibitem{Belanger:2013pna}
G.~B\'elanger, G.~Drieu La~Rochelle, B.~Dumont, R.~M. Godbole, S.~Kraml, and
  S.~Kulkarni, {\it {LHC constraints on light neutralino dark matter in the
  MSSM}},  {\em Phys. Lett. B} {\bf 726} (2013) 773--780,
  [\href{http://arxiv.org/abs/1308.3735}{{\tt arXiv:1308.3735}}].

\bibitem{Bechtle:2015nua}
P.~Bechtle et~al., {\it {Killing the cMSSM softly}},  {\em Eur. Phys. J. C}
  {\bf 76} (2016), no.~2 96, [\href{http://arxiv.org/abs/1508.05951}{{\tt
  arXiv:1508.05951}}].

\bibitem{deVries:2015mfw}
J.~de~Vries, H.~K. Dreiner, and D.~Schmeier, {\it {R-Parity Violation and Light
  Neutralinos at SHiP and the LHC}},  {\em Phys. Rev. D} {\bf 94} (2016), no.~3
  035006, [\href{http://arxiv.org/abs/1511.07436}{{\tt arXiv:1511.07436}}].

\bibitem{Wang:2019orr}
Z.~S. Wang and K.~Wang, {\it {Long-lived light neutralinos at future
  $Z-$factories}},  {\em Phys. Rev. D} {\bf 101} (2020), no.~11 115018,
  [\href{http://arxiv.org/abs/1904.10661}{{\tt arXiv:1904.10661}}].

\bibitem{Wang:2019xvx}
Z.~S. Wang and K.~Wang, {\it {Physics with far detectors at future lepton
  colliders}},  {\em Phys. Rev. D} {\bf 101} (2020), no.~7 075046,
  [\href{http://arxiv.org/abs/1911.06576}{{\tt arXiv:1911.06576}}].

\bibitem{Helo:2018qej}
J.~C. Helo, M.~Hirsch, and Z.~S. Wang, {\it {Heavy neutral fermions at the
  high-luminosity LHC}},  {\em JHEP} {\bf 07} (2018) 056,
  [\href{http://arxiv.org/abs/1803.02212}{{\tt arXiv:1803.02212}}].

\bibitem{Dercks:2018eua}
D.~Dercks, J.~De~Vries, H.~K. Dreiner, and Z.~S. Wang, {\it {R-parity Violation
  and Light Neutralinos at CODEX-b, FASER, and MATHUSLA}},  {\em Phys. Rev. D}
  {\bf 99} (2019), no.~5 055039, [\href{http://arxiv.org/abs/1810.03617}{{\tt
  arXiv:1810.03617}}].

\bibitem{Dercks:2018wum}
D.~Dercks, H.~K. Dreiner, M.~Hirsch, and Z.~S. Wang, {\it {Long-Lived Fermions
  at AL3X}},  {\em Phys. Rev. D} {\bf 99} (2019), no.~5 055020,
  [\href{http://arxiv.org/abs/1811.01995}{{\tt arXiv:1811.01995}}].

\bibitem{Dreiner:2020qbi}
H.~K. Dreiner, J.~Y. G\"unther, and Z.~S. Wang, {\it {R-parity Violation and
  Light Neutralinos at ANUBIS and MAPP}},
  \href{http://arxiv.org/abs/2008.07539}{{\tt arXiv:2008.07539}}.

\bibitem{Achasov:2019rdp}
{\bf BES-III} Collaboration, M.~Achasov, X.~Mo, N.~Muchnoi, I.~Nikolaev,
  S.~Privalov, and J.~Zhang, {\it {Tau mass measurement at BES-III}},  {\em EPJ
  Web Conf.} {\bf 212} (2019) 08005.

\bibitem{Abe:2010gxa}
{\bf Belle-II} Collaboration, T.~Abe et~al., {\it {Belle II Technical Design
  Report}},  \href{http://arxiv.org/abs/1011.0352}{{\tt arXiv:1011.0352}}.

\bibitem{Kou:2018nap}
{\bf Belle-II} Collaboration, W.~Altmannshofer et~al., {\it {The Belle II
  Physics Book}},  {\em PTEP} {\bf 2019} (2019), no.~12 123C01,
  [\href{http://arxiv.org/abs/1808.10567}{{\tt arXiv:1808.10567}}]. [Erratum:
  PTEP 2020, 029201 (2020)].

\bibitem{Aubert:2001tu}
{\bf BaBar} Collaboration, B.~Aubert et~al., {\it {The BaBar detector}},  {\em
  Nucl. Instrum. Meth. A} {\bf 479} (2002) 1--116,
  [\href{http://arxiv.org/abs/hep-ex/0105044}{{\tt hep-ex/0105044}}].

\bibitem{Abashian:2000cg}
A.~Abashian et~al., {\it {The Belle Detector}},  {\em Nucl. Instrum. Meth. A}
  {\bf 479} (2002) 117--232.

\bibitem{Batell:2009yf}
B.~Batell, M.~Pospelov, and A.~Ritz, {\it {Probing a Secluded U(1) at
  B-factories}},  {\em Phys. Rev. D} {\bf 79} (2009) 115008,
  [\href{http://arxiv.org/abs/0903.0363}{{\tt arXiv:0903.0363}}].

\bibitem{Canetti:2014dka}
L.~Canetti, M.~Drewes, and B.~Garbrecht, {\it {Probing leptogenesis with
  GeV-scale sterile neutrinos at LHCb and Belle II}},  {\em Phys. Rev. D} {\bf
  90} (2014), no.~12 125005, [\href{http://arxiv.org/abs/1404.7114}{{\tt
  arXiv:1404.7114}}].

\bibitem{Dolan:2017osp}
M.~J. Dolan, T.~Ferber, C.~Hearty, F.~Kahlhoefer, and K.~Schmidt-Hoberg, {\it
  {Revised constraints and Belle II sensitivity for visible and invisible
  axion-like particles}},  {\em JHEP} {\bf 12} (2017) 094,
  [\href{http://arxiv.org/abs/1709.00009}{{\tt arXiv:1709.00009}}].

\bibitem{Sullivan:2017qdj}
M.~K. Sullivan and D.~Fryberger, {\it {Magnetic Charge Search for the BELLE II
  Detector}},  \href{http://arxiv.org/abs/1707.05295}{{\tt arXiv:1707.05295}}.

\bibitem{Cvetic:2017vwl}
G.~Cvetic and C.~Kim, {\it {Sensitivity limits on heavy-light mixing $|U_{\mu
  N}|^2$ from lepton number violating $B$ meson decays}},  {\em Phys. Rev. D}
  {\bf 96} (2017), no.~3 035025, [\href{http://arxiv.org/abs/1705.09403}{{\tt
  arXiv:1705.09403}}]. [Erratum: Phys.Rev.D 102, 019903 (2020), Erratum:
  Phys.Rev.D 102, 039902 (2020)].

\bibitem{Dib:2019tuj}
C.~Dib, J.~Helo, M.~Nayak, N.~Neill, A.~Soffer, and J.~Zamora-Saa, {\it
  {Searching for a sterile neutrino that mixes predominantly with $\nu_\tau$ at
  $B$ factories}},  {\em Phys. Rev. D} {\bf 101} (2020), no.~9 093003,
  [\href{http://arxiv.org/abs/1908.09719}{{\tt arXiv:1908.09719}}].

\bibitem{Filimonova:2019tuy}
A.~Filimonova, R.~Sch\"afer, and S.~Westhoff, {\it {Probing dark sectors with
  long-lived particles at BELLE II}},  {\em Phys. Rev. D} {\bf 101} (2020),
  no.~9 095006, [\href{http://arxiv.org/abs/1911.03490}{{\tt
  arXiv:1911.03490}}].

\bibitem{Duerr:2019dmv}
M.~Duerr, T.~Ferber, C.~Hearty, F.~Kahlhoefer, K.~Schmidt-Hoberg, and
  P.~Tunney, {\it {Invisible and displaced dark matter signatures at Belle
  II}},  {\em JHEP} {\bf 02} (2020) 039,
  [\href{http://arxiv.org/abs/1911.03176}{{\tt arXiv:1911.03176}}].

\bibitem{SHiP:2018yqc}
{\bf SHiP} Collaboration, C.~Ahdida et~al., {\it {The experimental facility for
  the Search for Hidden Particles at the CERN SPS}},  {\em JINST} {\bf 14}
  (2019), no.~03 P03025, [\href{http://arxiv.org/abs/1810.06880}{{\tt
  arXiv:1810.06880}}].

\bibitem{Bansal:2019zak}
S.~Bansal, A.~Delgado, C.~Kolda, and M.~Quiros, {\it {Constraining
  $R$-parity-violating couplings in $\tau$-processes at the LHC and in
  electroweak precision measurements}},  {\em Phys. Rev. D} {\bf 100} (2019),
  no.~9 093005, [\href{http://arxiv.org/abs/1906.01063}{{\tt
  arXiv:1906.01063}}].

\bibitem{Aaboud:2018vgh}
{\bf ATLAS} Collaboration, M.~Aaboud et~al., {\it {Search for High-Mass
  Resonances Decaying to $\tau\nu$ in pp Collisions at $\sqrt{s}$=13 TeV with
  the ATLAS Detector}},  {\em Phys. Rev. Lett.} {\bf 120} (2018), no.~16
  161802, [\href{http://arxiv.org/abs/1801.06992}{{\tt arXiv:1801.06992}}].

\bibitem{1807.11421}
{\bf CMS} Collaboration, A.~M. Sirunyan et~al., {\it {Search for a W' boson
  decaying to a $\tau$ lepton and a neutrino in proton-proton collisions at
  $\sqrt{s} =$ 13 TeV}},  {\em Phys. Lett. B} {\bf 792} (2019) 107--131,
  [\href{http://arxiv.org/abs/1807.11421}{{\tt arXiv:1807.11421}}].

\bibitem{Zyla:2020zbs}
{\bf Particle Data Group} Collaboration, P.~Zyla et~al., {\it {Review of
  Particle Physics}},  {\em PTEP} {\bf 2020} (2020), no.~8 083C01.

\bibitem{Decker:1994kw}
R.~Decker and M.~Finkemeier, {\it {Radiative corrections to the decay tau
  ---\ensuremath{>} pi tau-neutrino}},  {\em Nucl. Phys. B Proc. Suppl.} {\bf
  40} (1995) 453--461, [\href{http://arxiv.org/abs/hep-ph/9411316}{{\tt
  hep-ph/9411316}}].

\bibitem{delAmoSanchez:2010pc}
{\bf BaBar} Collaboration, P.~del Amo~Sanchez et~al., {\it {Studies of tau-
  ---\ensuremath{>} eta K-nu and tau- ---\ensuremath{>} eta pi- nu(tau) at
  BaBar and a search for a second-class current}},  {\em Phys. Rev. D} {\bf 83}
  (2011) 032002, [\href{http://arxiv.org/abs/1011.3917}{{\tt
  arXiv:1011.3917}}].

\bibitem{Pich:2013lsa}
A.~Pich, {\it {Precision Tau Physics}},  {\em Prog. Part. Nucl. Phys.} {\bf 75}
  (2014) 41--85, [\href{http://arxiv.org/abs/1310.7922}{{\tt
  arXiv:1310.7922}}].

\bibitem{Nussinov:2009sn}
S.~Nussinov and A.~Soffer, {\it {Estimate of the Branching Fraction of $\tau
  \to \pi \eta' \nu_\tau$}},  {\em Phys. Rev. D} {\bf 80} (2009) 033010,
  [\href{http://arxiv.org/abs/0907.3628}{{\tt arXiv:0907.3628}}].

\bibitem{Nussinov:2008gx}
S.~Nussinov and A.~Soffer, {\it {Estimate of the branching fraction tau
  ---\ensuremath{>} eta pi- nu(tau), the a(0)-(980), and non-standard weak
  interactions}},  {\em Phys. Rev. D} {\bf 78} (2008) 033006,
  [\href{http://arxiv.org/abs/0806.3922}{{\tt arXiv:0806.3922}}].

\bibitem{Lange:2001uf}
D.~Lange, {\it {The EvtGen particle decay simulation package}},  {\em Nucl.
  Instrum. Meth. A} {\bf 462} (2001) 152--155.

\bibitem{Allmendinger:2012ch}
T.~Allmendinger et~al., {\it {Track Finding Efficiency in BaBar}},  {\em Nucl.
  Instrum. Meth. A} {\bf 704} (2013) 44--59,
  [\href{http://arxiv.org/abs/1207.2849}{{\tt arXiv:1207.2849}}].

\bibitem{TheBABAR:2013jta}
{\bf BaBar} Collaboration, B.~Aubert et~al., {\it {The BABAR Detector:
  Upgrades, Operation and Performance}},  {\em Nucl. Instrum. Meth. A} {\bf
  729} (2013) 615--701, [\href{http://arxiv.org/abs/1305.3560}{{\tt
  arXiv:1305.3560}}].

\bibitem{Sjostrand:2006za}
T.~Sjostrand, S.~Mrenna, and P.~Z. Skands, {\it {PYTHIA 6.4 Physics and
  Manual}},  {\em JHEP} {\bf 05} (2006) 026,
  [\href{http://arxiv.org/abs/hep-ph/0603175}{{\tt hep-ph/0603175}}].

\bibitem{Sjostrand:2007gs}
T.~Sjostrand, S.~Mrenna, and P.~Z. Skands, {\it {A Brief Introduction to PYTHIA
  8.1}},  {\em Comput. Phys. Commun.} {\bf 178} (2008) 852--867,
  [\href{http://arxiv.org/abs/0710.3820}{{\tt arXiv:0710.3820}}].

\bibitem{0905.2051}
H.~Dreiner, S.~Grab, D.~Koschade, M.~Kramer, B.~O'Leary, and U.~Langenfeld,
  {\it {Rare meson decays into very light neutralinos}},  {\em Phys. Rev. D}
  {\bf 80} (2009) 035018, [\href{http://arxiv.org/abs/0905.2051}{{\tt
  arXiv:0905.2051}}].

\bibitem{Bloch:1999vka}
J.~C. Bloch, Y.~Kalinovsky, C.~D. Roberts, and S.~Schmidt, {\it {Describing
  a(1) and b(1) decays}},  {\em Phys. Rev. D} {\bf 60} (1999) 111502,
  [\href{http://arxiv.org/abs/nucl-th/9906038}{{\tt nucl-th/9906038}}].

\bibitem{Cheng:2003bn}
H.-Y. Cheng, {\it {Hadronic charmed meson decays involving axial vector
  mesons}},  {\em Phys. Rev. D} {\bf 67} (2003) 094007,
  [\href{http://arxiv.org/abs/hep-ph/0301198}{{\tt hep-ph/0301198}}].

\bibitem{Ebert:2006hj}
D.~Ebert, R.~Faustov, and V.~Galkin, {\it {Relativistic treatment of the decay
  constants of light and heavy mesons}},  {\em Phys. Lett. B} {\bf 635} (2006)
  93--99, [\href{http://arxiv.org/abs/hep-ph/0602110}{{\tt hep-ph/0602110}}].

\bibitem{Dreiner:2006gu}
H.~Dreiner, M.~Kramer, and B.~O'Leary, {\it {Bounds on R-parity violating
  supersymmetric couplings from leptonic and semi-leptonic meson decays}},
  {\em Phys. Rev. D} {\bf 75} (2007) 114016,
  [\href{http://arxiv.org/abs/hep-ph/0612278}{{\tt hep-ph/0612278}}].

\end{thebibliography}\endgroup

\end{document}